\documentclass[aps,prd,showpacs,superscriptaddress,notitlepage,nofootinbib]{revtex4-1}
\usepackage{graphicx}  
\usepackage{dcolumn}   
\usepackage{bm}        
\usepackage{amssymb}   
\usepackage{amsmath,amssymb,bm,graphicx,bbold,epsf,colordvi}
\usepackage{dsfont}
\usepackage{color}
\def\d{{ {d}}}
\def\OMIT#1{{}}

\newcommand{\ba}{\begin{eqnarray}}
\newcommand{\ea}{\end{eqnarray}}

\newcommand{\be}{\begin{equation}}
\newcommand{\ee}{\end{equation}}
\newcommand{\bea}{\begin{eqnarray}}
\newcommand{\eea}{\end{eqnarray}}

\def\beaa#1\eeaa{\begin{align}#1\end{align}} 
\usepackage{amsmath}
\def\d{{\rm d}}
\def\OMIT#1{{}}
\newcommand{\mcdot}{\!\cdot\!}

\newcommand{\eq}[1]{Eq.~\eqref{#1}}
\newcommand{\vc}[1]{\boldsymbol{#1}}

\begin{document}

\title{Jet Quenching from QCD Evolution}

\author{Yang-Ting Chien} 
\email{ytchien@lanl.gov}
\affiliation{Theoretical Division, Los Alamos National Laboratory, Los Alamos, NM 87545, USA}

\author{Alexander Emerman}
\email{aze2001@columbia.edu}
\affiliation{Department of Physics, Columbia University, New York, NY 10027, USA}

\author{Zhong-Bo Kang}
\email{zkang@lanl.gov}
\affiliation{Theoretical Division, Los Alamos National Laboratory, Los Alamos, NM 87545, USA}

\author{Grigory Ovanesyan} 
\email{ovanesyan@umass.edu}
\affiliation{Physics Department, University of Massachusetts Amherst, Amherst, MA 01003, USA}

\author{Ivan Vitev} 
\email{ivitev@lanl.gov}
\affiliation{Theoretical Division, Los Alamos National Laboratory, Los Alamos, NM 87545, USA}

\date{\today}

\begin{abstract}
Recent advances in soft-collinear effective theory with Glauber gluons have led to the development of a new method that gives a unified description of inclusive hadron production in reactions with nucleons and heavy nuclei. We show how this approach,  based on  the generalization of the DGLAP evolution equations to include final-state medium-induced parton shower corrections for large $Q^2$ processes,  can be combined with initial-state effects for applications to jet quenching phenomenology. We demonstrate that the traditional parton energy loss calculations can be regarded as a special soft-gluon emission limit of the general QCD evolution framework.  We present phenomenological comparison of the SCET$_{\rm G}$-based results on the suppression of inclusive charged hadron and neutral pion production in $\sqrt{s_{NN}}=2.76$~TeV lead-lead collisions at the Large Hadron Collider to experimental data.  We also show theoretical predictions  for the upcoming  $\sqrt{s_{NN}} \simeq 5.1$~TeV Pb+Pb run at the LHC.
\end{abstract}

\preprint{ACFI-T15-13}
\maketitle

\section{Introduction}\label{sec:introduction}

Understanding parton shower formation and evolution is central to perturbative Quantum Chromodynamics (pQCD)~\cite{Brock:1993sz}. Parton showers connect the short-distance physics of hard, large $Q^2$, scattering  with the long-distance non-perturbative dynamics of hadronization, govern the formations of jets~\cite{Sterman:1977wj,Ellis:1990ek}, and control the
evolution of parton distribution and fragmentation functions through the standard Dokshitzer-Gribov-Lipatov-Altarelli-Parisi  (DGLAP) evolution equations~\cite{Gribov:1972ri,Dokshitzer:1977sg,Altarelli:1977zs}.
 High energy scattering processes naturally present a  multi-scale problem, ideally suited to effective field theory treatment. Indeed, over the past decade many of the advances in understanding parton shower formation and the resummation of large logarithms that arise from ratios of energy and momentum scales in e$^+$+e$^-$, e+p and p+p have come from  the well-established soft-collinear effective theory (SCET)~\cite{Bauer:2000ew,Bauer:2000yr,Bauer:2001ct,Bauer:2001yt,Becher:2014oda}, an effective theory of QCD for jet physics.  Recently, SCET has been extended to describe jet propagation in  matter, where the physical interactions with the medium are mediated via Glauber gluon exchange~\cite{Idilbi:2008vm,DEramo:2010ak,Ovanesyan:2011xy}.
The resulting effective theory SCET$_{\rm G}$ has been used to derive all  $O(\alpha_s)$ $1\to 2$ medium-induced splitting kernels \cite{Ovanesyan:2011kn} and discuss higher order $O(\alpha_s^2 )$ corrections to the  medium-modified jet substructure~\cite{Fickinger:2013xwa}.

Such advances in the theory of in-medium parton shower formation have allowed us to overcome some of the inherent limitations of the traditional energy loss approach in nucleus-nucleus (A+A) collisions, see for example~\cite{Gyulassy:2003mc}, and to unify our understanding of energetic particle and jet production in p+p and A+A~\cite{Kang:2014xsa}.  In the current paper, we provide the details of  the implementation of the in-medium QCD evolution-based framework and give an extended discussions of the connection between the energy loss approach and our new method for evaluating hadron production in the soft gluon emission limit. To address charged hadron and neutral pion production in Pb+Pb reactions at $\sqrt{s_{NN}}=2.76$~TeV, we combine the medium-modified fragmentation functions with initial-state cold nuclear matter effects. We find that this theoretical model gives a good description of ALICE, ATLAS and CMS experimental 
measurements~\cite{Abelev:2012hxa,Abelev:2014ypa,Aad:2015wga,CMSraa}  
of jet quenching, the attenuation in the  production rate of energetic particles and jets in heavy ion reactions relative to the p+p baseline scaled by the number of elementary nucleon-nucleon interactions introduced in~\cite{Wang:1991xy}.  While limited theoretical predictions for open heavy flavor at higher center-of-mass energies have been made available~\cite{Djordjevic:2015hra}, results on light hadron production are generally absent from the literature.
We take this opportunity to present theoretical predictions for the anticipated Pb+Pb run at $\sqrt{s_{NN}} \simeq 5.1$~TeV at the LHC. We note that the in-medium evolution approach has been previously applied to semi-inclusive
deep inelastic scattering~\cite{Wang:2009qb,Chang:2014fba}.

The rest of our paper is organized as follows. In Sec.~\ref{sec:energy-loss}, we discuss the strengths and limitations of the traditional energy loss  approach to facilitate our comparison with the new QCD evolution technique. In Sec.~\ref{sec:Splitting}, we provide  details on how to derive the full medium-induced splitting functions and fix the virtual correction pieces through flavor and momentum sum rules. We demonstrate the connection between the energy loss approach and the QCD evolution framework  in the soft gluon approximation. In Sec.~\ref{sec:hadron}, we combine the final-state in-medium parton shower evolution with initial-state effects to provide detailed comparison to the experimental data on  inclusive hadron production at the LHC. We conclude our paper in Sec.~\ref{sec:conclusions}. The full result for the medium-induced splitting kernels,  some details about the phenomenological implementation and numerical evaluation of these kernels, and details about the implementation of cold nuclear matter effects are collected in the Appendix.

\section{The energy loss-based approach}
\label{sec:energy-loss}

High transverse momentum production of energetic particles in relativistic heavy ion collisions has played an essential role in probing the properties of the quark-gluon plasma  (QGP)  at the Relativistic Heavy Ion Collider (RHIC) and the Large Hadron Collider (LHC). It was realized early on that as energetic quarks and gluons propagate through the QGP, they will interact with the medium and lose energy via collisional and radiative processes. The related attenuation of  particle flux, commonly referred to as jet quenching~\cite{Wang:1991xy,Gyulassy:1993hr} has attracted tremendous attention from both the theoretical and experimental communities. There has been great progress in studying jet quenching phenomena based on the parton energy loss picture, ranging from  inclusive light hadron and heavy meson  production~\cite{Adcox:2001jp,Adler:2002xw,Vitev:2002pf,Djordjevic:2004nq,Aamodt:2010jd,Abelev:2012hxa,Aad:2015wga,Adam:2015kca,Abelev:2014ypa,CMS:2012aa,ALICE:2012ab,Sharma:2009hn} to di-hadron correlations~\cite{Adler:2002tq,Renk:2006pk,Adare:2010ry,Zhang:2007ja} and $\gamma$+hadron correlations~\cite{Adare:2009vd, Abelev:2009gu,Zhang:2009rn,Kang:2011rt}.

An important step forward  in understanding hard processes in the presence of a QCD medium was to realize that reconstructed jets will also be modified~\cite{Salgado:2003rv} and to develop a theory that describes 
and connects jet cross section and
jet substructure related observables in heavy ion reactions~\cite{Vitev:2008rz}.  
Theoretical predictions compared to experimental measurements of inclusive jets~\cite{Vitev:2009rd,He:2011pd,Aad:2012vca},   
di-jets~\cite{Aad:2010bu,Chatrchyan:2011sx,Qin:2010mn,He:2011pd},  $Z^0/\gamma$-tagged jets~\cite{Neufeld:2010fj,Chatrchyan:2012gt,Dai:2012am,Ma:2013bia} and  heavy flavor jets production~\cite{Chatrchyan:2013exa,Huang:2013vaa,Huang:2015mva} have allowed characterization of the angular and momentum distributions of the soft medium-induced gluon bremsstrahlung. 
 One approximation that is inherent in radiative energy loss-based predictions is that the fractional energy loss for 
the parent partons per emitted gluon is small and, thus, the parent partons do not change their identity. 
This limit is usually referred to as the soft gluon limit (or small-$x$ limit, where $x=k^+/p_0^+$ is the 
light-cone momentum fraction carried away by the gluon). Large energy loss in this case proceeds through
multiple gluon emission.  Large energy loss can also result from 
collisional drag-like processes for the in-medium parton shower~\cite{Neufeld:2011yh,Casalderrey-Solana:2014bpa} 
both in the weak and the  strong coupling limits.

Two approaches to parton energy loss, the High Twist (HT) approach~\cite{Wang:2001ifa,Wang:2002ri} and the 
Guylassy-Levai-Vitev (GLV) approach~\cite{Gyulassy:2000er,Gyulassy:2000fs} generalized to massive partons by Djordjevic~\cite{Djordjevic:2003zk}, start with the picture of factorization of hard processes in 
QCD and treat the medium-induced splittings as process-dependent radiative corrections to the factorized 
expressions  for hadronic, leptonic and jet observables.   
 This makes it conceptually simpler to generalize the soft gluon emission limit to full parton shower treatment.  
We briefly discuss  the energy loss approach on the example of the GLV solution for the medium-induced radiative spectrum. To the first order in opacity~\cite{Vitev:2007ve}, the distribution of inclusive gluon radiation is given by
\beaa
x\frac{d^3N_g}{dx d^2 \vc{k}_\perp}   = &\frac{C_{R} \alpha_s}{\pi^2} 
 \int_0^L  \frac{d \Delta z}{\lambda_{g} } 
\int  d^2 \vc{q}_\perp  \frac{1}{\sigma_{el}}    \frac{d^2\sigma_{el}^{\rm medium}}{d^2 \vc{q}_\perp}  
\frac{ 2 \vc{k}_\perp \cdot \vc{q}_\perp }{\vc{k}_\perp^2 (\vc{k}_\perp-\vc{q}_\perp)^2}   
\left[ 1 - \cos\left(   \frac{ (\vc{k}_\perp-\vc{q}_\perp)^2}{x p_0^+} \Delta z \right) \right],
\label{double}
\eeaa
where $C_R$ is the quadratic Casimir in the fundamental or adjoint representation of SU(3) when the parent parton is a quark or a gluon, respectively. In Eq.~(\ref{double}) $\lambda_g$ is the gluon scattering length of ${\mathcal O}$(1 fm),  $x=k^+/p_0^+\approx \omega/E$ is the fractional energy of parent parton carried by the emitted gluon, $\vc{k}_\perp$ is its momentum transverse to the jet axis while $\vc{q}_\perp$ is the transverse momentum exchange between the propagating parton and the QGP, and $\frac{1}{\sigma_{el}}    \frac{d^2\sigma_{el}^{\rm medium}}{d^2 \vc{q}_\perp}$ is the normalized differential distribution of the elastic scattering. The details of evaluating such a double differential distribution in Eq.~\eqref{double} are given in, e.g. Refs.~\cite{Vitev:2007ve,Vitev:2008rz,Vitev:2009rd}. From this distribution one can obtain the gluon number/intensity spectra, as well as the average gluon number/parton energy loss:
\beaa
\frac{dN_g}{dx}  = \int d^2 \vc{k}_{\perp}  \frac{d^3N_g}{dx d^2\vc{k}_{\perp}}, 
\qquad
\langle N_g\rangle  =  \int d x  \,  \frac{dN_g}{dx}   \, , 
\qquad       
\langle \Delta E \rangle = \int dx   \,  xE\, \frac{dN_g}{dx}\,.
\eeaa
There has also been recent interest in the scale dependence of the transport properties of matter probed in scattering and energy loss processes~\cite{Kang:2013raa}.

Despite the success of jet quenching phenomenology described above, it is now  clear that qualitative advances in
understanding QCD in the heavy ion environment are needed. Radiative energy loss calculations cannot be systematically improved because this approach is well defined only in the soft gluon emission limit when the parent parton does not change its identity. Thus, the concept of a higher order energy loss calculation is not well defined.  Furthermore,  for quarks and gluons to lose a sizable fraction of their energy would require the process to take place through multiple gluon 
emission. Precluding hard splitting processes can significantly affect the accuracy of the theoretical predictions for 
di-hadron correlations, tagged jets and di-jets~\cite{Kang:2014xsa}. Detailed understanding of in-medium showers is
also a prerequisite for incorporating collisional energy loss. Most importantly, the energy loss approach cannot incorporate the advances in  understanding higher order calculations and
resummation in the standard pQCD and SCET frameworks.

Given the success of parton energy loss phenomenology, it is important to also validate this vast body of existing 
work and demonstrate the connection to the general QCD evolution framework in the presence of a medium. In Ref.~\cite{Kang:2014xsa}   
it was shown numerically in great detail that for sufficiently inclusive observables, such as high $p_T$
hadron production, the two approaches give practically identical results when multiple gluon emission is 
considered~\cite{Vitev:2005he}. Any differences can be re-absorbed in a $\lesssim 5\%$ change of the coupling $g$ between the 
jet and the medium. We demonstrate below that the energy loss approach is related to a special soft gluon  limit solution to the DGLAP evolution equations with full splitting functions,
when the medium-induced corrections are included.


\section{Splitting functions and QCD evolution of fragmentation functions}
\label{sec:Splitting}
In this section we review the splitting functions in both the vacuum and a QCD medium. These govern the DGLAP evolution equations of fragmentation functions which encode the effect of multiple gluon emissions. We will show that this QCD-evolution based formulation is connected to the energy loss approximation in the soft-gluon limit. Since the evolution equations do not make this approximation, they consistently take into account the full-$x$ effects in the medium, including flavor changing processes $g \leftrightarrow q,{\bar{q}}$.

\subsection{Splitting functions and DGLAP evolution in the vacuum}
\label{sec:vacuumsplittingfunctions}
First, we will review the leading order splitting functions in the vacuum. These are well-known and reproduced using the SCET collinear Lagrangian \cite{Baumgart:2010qf,Ovanesyan:2011kn}. The real emission contribution of the splitting kernels are calculated from the tree level diagrams of $1\rightarrow 2$ splitting. In the splitting $a\rightarrow bc$ we define $x$ to be the momentum fraction of the second final state parton $c$. For example, in the splitting $q\rightarrow qg$, the gluon carries the fraction $x$ of the parent quark momentum. This is opposite,  $x \leftrightarrow 1-x $, to traditional notation but ensures that $x\ll 1$ corresponds to the soft gluon emission limit.  With this convention, the $q\rightarrow qg$ splitting kernel reads:
\begin{eqnarray}
\label{qqg}
\frac{dN^{(i)}_{\rm vac}}{dx \,d^2{\vc{k}}}_{\perp}&=&
\frac{\alpha_s}{2\pi^2} C_F \frac{1+(1-x)^2}{x}\frac{1}{{\vc{k}}_{\perp}^2}\equiv \frac{\alpha_s}{2\pi^2} \frac{1}{{\vc{k}}_{\perp}^2} P^{\text{vac,~real}}_{qq}(x),
 \end{eqnarray}
where $(i)={q\rightarrow qg}$ and the subscript ``vac'' represents the splitting functions in the vacuum, and  
\beaa
P^{\text{vac,~real}}_{qq}(x) = C_F\frac{1+(1-x)^2}{x}\, ,
\eeaa
is the corresponding leading-order splitting function. Similarly,
\begin{eqnarray}
&&P^{\text{vac,~real}}_{gg}(x)=2C_A\left(\frac{1-x}{x}+\frac{x}{1-x}+x(1-x)\right),\\
&&P^{\text{vac,~real}}_{gq}(x)=T_R\left(x^2+(1-x)^2\right),\\
&&P^{\text{vac,~real}}_{qg}(x)=C_F\frac{1+x^2}{1-x}.
\end{eqnarray}
Note that the splitting functions are labeled as $P_{ab}^{\rm vac,~real}(x)$ for $a\rightarrow bc$ splittings. In the energy loss approximation, the parent parton $a$ emits away the soft parton $c$, and the parton $b$ subsequently fragments into hadrons.

To include the virtual contributions, we use the flavor and momentum sum rules~\cite{Altarelli:1977zs} to fix the coefficients of the $\delta$-function pieces in the splitting functions. The splitting functions combining real and virtual contributions take the following forms,
\begin{eqnarray}
&&P_{qq}^{\rm vac}(x)=\left[P^{\text{vac,~real}}_{qq}(x)\right]_++A\,\delta(x),\\
&&P_{gg}^{\rm vac}(x)=2C_A\left(\left[\frac{1-2x}{x}+x(1-x)\right]_++\frac{1}{1-x}\right)+B\,\delta(x),\\
&&P_{gq}^{\rm vac}(x)=P^{\text{vac,~real}}_{gq}(x),\\[1.5ex]
&&P_{qg}^{\rm vac}(x)=P^{\text{vac,~real}}_{qg}(x),
\end{eqnarray}
where we have written the real emission contributions as plus functions. The last two splitting functions do not have $\delta$-function pieces since there does not exist a one-loop virtual diagram which would change the flavor of the parton. The flavor and momentum conservation sum rules are given by,
\begin{eqnarray}
&&\int_0^1 P_{qq}^{\rm vac}(x)\,\d x=0,\\
&&\int_0^1 \left[P_{qq}^{\rm vac}(x)+P_{qg}^{\rm vac}(x)\right](1-x)\,\d x=0,\\
&&\int_0^1 \left[2n_fP_{gq}^{\rm vac}(x)+P_{gg}^{\rm vac}(x)\right](1-x)\,\d x=0,
\end{eqnarray}
with $n_f$ the number of active quark flavors. Because $P^{\text{vac,~real}}_{qq}$ and $P^{\text{vac,~real}}_{qg}$ are related by $x\leftrightarrow 1-x$, the second sum rule follows from the first one. Therefore we obtain
\beaa
A=&0,\\
B=&\int\d x'\left\{2C_A\left(x'\left(\frac{1-2x'}{x'}+x'(1-x')\right)-1\right)-2n_f(1-x')P^{\rm vac}_{g\rightarrow q\bar{q}}(x')\right\}\nonumber\\
=&-\frac{11 C_A}{6}-\frac{2 n_f T_R}{3}.
\eeaa
Note that our plus function pieces are defined slightly differently than those in the literature. In particular, we combine some extra $x$-dependent terms into the plus functions and there is a difference between $\left[f(x)/x\right]_+$ and $f(x)/(x)_+$. In the small-$x$ approximation, the real contributions to the splitting functions reduce to:
\begin{eqnarray}
&&P^{\text{vac,~real}}_{qq}(x)\approx \frac{2\,C_F}{x}, \qquad P^{\text{vac,~real}}_{gg}(x)\approx \frac{2\,C_A}{x},\\
&&P^{\text{vac,~real}}_{gq}(x)\approx 0,\qquad   P^{\text{vac,~real}}_{qg}(x)\approx 0.
\end{eqnarray}
The flavor-diagonal splitting functions dominate, and the parton does not change its flavor in this limit.

The fragmentation functions satisfy the DGLAP evolution equations,
\beaa
\frac{\d D_{h/q}(z,Q)}{\d \ln Q}&=\frac{\alpha_s(Q)}{\pi}\int_{z}^1 \frac{\d z'}{z'}\left[P^{\rm vac}_{q\rightarrow qg}(z')D_{h/q}\left(\frac{z}{z'},Q\right)+P^{\rm vac}_{q\rightarrow gq}(z') D_{h/g}\left(\frac{z}{z'},Q\right)\right],
\label{eq:AP10}
\\
\frac{\d D_{h/g}(z,Q)}{\d \ln Q}&=\frac{\alpha_s(Q)}{\pi}\int_{z}^1 \frac{\d z'}{z'}\left[P^{\rm vac}_{g\rightarrow gg}(z')D_{h/g}\left(\frac{z}{z'},Q\right)+P^{\rm vac}_{g\rightarrow q\bar q}(z')\sum_q D_{h/q}\left(\frac{z}{z'},Q\right)\right],
\label{eq:AP30}
\eeaa
where we have relabeled the splitting functions with $|\vc{k}_\perp| = Q$ and $z= 1-x$. It is instructive to emphasize that the $\ln Q$-enhancement in the above DGLAP evolution equations can be traced back to the $1/\vc{k}_{\perp}^2$ behavior in $\frac{dN}{dx \,d^2{\vc{k}}_{\perp}}$ over a broad phase space in Q, see Eq.~\eqref{qqg}. We will comment how such a connection changes in the medium case. 
The evolved fragmentation functions can be used to describe the inclusive hadron transverse momentum spectrum in p+p collisions. In the next subsection we will follow a similar procedure to describe fragmentation function evolution in A+A collisions.

\subsection{Medium-induced splitting functions and generalized DGLAP evolution}
\label{sec:mediuminducedfunctions}
The real contributions of the medium-induced splitting kernels have been computed using SCET$_{\rm G}$ \cite{Ovanesyan:2011kn} to the first order in opacity. Quark-gluon splitting dressed with multiple soft interactions
beyond the soft emission limit has also been considered~\cite{Blaizot:2012fh,Apolinario:2014csa}.
Following the same definition as in Eq.~\eqref{qqg}, we have
\beaa
    \frac{\d N_{\rm med}^{(i)}}{\d x\,\d^2\vc{k}_{\perp}} = \frac{\alpha_s}{2\pi^2} \frac{1}{{\vc{k}}_{\perp}^2} P^{\text{med,~real}}_{i}(x,\vc{k}_{\perp})\, ,
\eeaa
where the subscript ``med'' emphasizes the medium-specific contribution, and $i=q\rightarrow qg,~q\rightarrow gg,~g\rightarrow q\bar{q}$ and $q\rightarrow gq$. Also, we define $h_{i}(x,\vc{k}_{\perp};\beta)$ as the ratio of the medium and the vacuum real emission splitting functions:
\begin{eqnarray}
&&P^{\text{med,~real}}_{q\rightarrow qg}(x, \vc{k}_{\perp}; \beta)= C_F\frac{1+(1-x)^2}{x}\,h_{q\to qg}\left(x,\vc{k}_{\perp}; \beta\right),\\
&&P^{\text{med,~real}}_{q\rightarrow gg}(x, \vc{k}_{\perp}; \beta)=2C_A\left(\frac{1-x}{x}+\frac{x}{1-x}+x(1-x)\right)\,h_{g\to gg}\left(x,\vc{k}_{\perp};\beta\right),\\
&&P^{\text{med,~real}}_{g\rightarrow q\bar{q}}(x, \vc{k}_{\perp}; \beta)=T_R\left(x^2+(1-x)^2\right)\,h_{g\to q\bar q}\left(x,\vc{k}_{\perp};\beta\right),\\[1ex]
&&P^{\text{med,~real}}_{q\rightarrow gq}(x, \vc{k}_{\perp}; \beta)=C_F\frac{1+x^2}{1-x}\,h_{q\to gq}\left(x,\vc{k}_{\perp};\beta\right).
\end{eqnarray}
Here the index $\beta$ represents the medium properties, such as the medium size and temperature, which enter the medium-induced splitting function calculations. We have included the results for $h_{i}(x,\vc{k}_{\perp};\beta)$ in Appendix~\ref{medsplitA}.

We expect the real plus virtual medium-induced splitting functions to have the following form,
\beaa
P^{\rm med}_{q\rightarrow qg}(x, \vc{k}_{\perp}; \beta)=&\left[P^{\text{med,~real}}_{q\rightarrow qg}(x,  \vc{k}_\perp; \beta)\right]_++A(\vc{k}_\perp; \beta)\,\delta(x),
\\
P^{\rm med}_{g\rightarrow gg}(x, \vc{k}_{\perp}; \beta)=&2C_A\left(\left[\left(\frac{1-2x}{x}+x(1-x)\right)h_{g\to gg}\left(x,\vc{k}_{\perp};\beta\right)\right]_+ 
+\frac{h_{g\to gg}\left(x, \vc{k}_{\perp}; \beta \right)}{1-x}\right)+B(\vc{k}_\perp; \beta)\,\delta(x),
\\
P^{\rm med}_{g\rightarrow q\bar{q}}(x, \vc{k}_{\perp}; \beta)=&P^{\text{med,~real}}_{g\rightarrow q\bar{q}}(x, \vc{k}_{\perp}; \beta),\\[1.5ex]
P^{\rm med}_{q\rightarrow gq}(x, \vc{k}_{\perp}; \beta)=&P^{\text{med,~real}}_{q\rightarrow gq}(x, \vc{k}_\perp; \beta).
\eeaa 
From momentum and flavor sum rules, the coefficients $A$ and $B$, which in this case are functions of $\vc{k}_{\perp}$ and medium properties $\beta$, are equal to,
\begin{eqnarray}
A(\vc{k}_\perp; \beta)&=&0\, ,
\\
B(\vc{k}_\perp; \beta)&=&\int\d x'\left\{2C_A\left(x'\left(\frac{1-2x'}{x'}+x'(1-x')\right)-1\right)h_{g\to gg}\left(x',\vc{k}_{\perp}; \beta\right)
\right.
-2n_f(1-x')P^{\rm med}_{g\rightarrow q\bar{q}}(x ', \vc{k}_{\perp}; \beta)\Big\}\, . \quad
\end{eqnarray}
The {\it full} splitting functions in the medium are equal to the sum of the vacuum and the medium-induced ones
\begin{eqnarray}
P_{i}^{\text{full}}(x,\vc{k}_{\perp};\beta)=P_{i}^{\text{vac}}(x)+P_{i}^{\rm med}(x,\vc{k}_{\perp};\beta)\,.
\label{full}
\label{gensum}
\end{eqnarray}
Note that whereas the vacuum splitting functions are functions of the momentum fraction $x$ alone, the medium-induced splitting functions depend on both $x$ and $\vc{k}_\perp$. 
Furthermore,  $P_{i}^{\rm med}(x,\vc{k}_{\perp};\beta) = P_{i}^{\rm med, \,1}(x,\vc{k}_{\perp};\beta) + P_{i}^{\rm med, \,2}(x,\vc{k}_{\perp};\beta) + \cdots $ can be calculated order-by-order in the correlation between the hard scattering and the parton interactions in the medium, also known as opacity expansion. For medium-induced branching, only  $P_{i}^{\rm med, \,1}(x,\vc{k}_{\perp};\beta)$ is known beyond the
soft gluon emission limit~\cite{Ovanesyan:2011kn} and we use this result in the paper.  
In the small-$x$ limit, the full splitting functions in the medium reduce to,
\begin{eqnarray}
&&P^{\rm full}_{q\rightarrow qg}(x,\vc{k}_\perp;\beta)=\frac{2C_F}{(x)_+}+\left(\frac{2C_F}{x}\,h^{\text{sga}}(x,\vc{k}_\perp;\beta)\right)_+,\label{eq:Pqqsmallx}\\
&&P^{\rm full}_{g\rightarrow gg}(x,\vc{k}_\perp;\beta)=\frac{2C_A}{(x)_+}+\left(\frac{2C_A}{x}\,h^{\text{sga}}(x,\vc{k}_\perp;\beta)\right)_+,\label{eq:Pggsmallx}\\
&&P^{\rm full}_{g\rightarrow q\bar{q}}(x,\vc{k}_\perp;\beta)=0,\\[1.5ex]
&&P^{\rm full}_{q\rightarrow gq}(x,\vc{k}_\perp;\beta)=0,\label{eq:Pgqsmallx}
\end{eqnarray}
where the function $h^{\text{sga}}(x,\vc{k}_\perp;\beta)$ \footnote{The superscript ``sga" stands for ``soft gluon approximation".} is the same for quark and gluon splitting and can be found in  Appendix~\ref{softA}.  

Different evolution equations have been discussed in the literature, the best known examples being 
Balitsky-Fadin-Kuraev-Lipatov 
(BFKL)~\cite{Kuraev:1977fs,Balitsky:1978ic}   (evolution in $x$)  and DGLAP (evolution  in virtuality or $|\vc{k}_{\perp}|$). In the case of large $Q^2$ processes, it has been shown~\cite{Nagy:2009re}  that the lowest order DGLAP evolution generically arises in parton showers. Evolution equations can be derived from the probabilistic interpretation of splitting processes inside such a  shower. This probabilistic interpretation remains the same in the presence of in-medium interactions, as long as all relevant interference terms for the final state interactions in the medium are included properly in the {\it full} splitting functions~\cite{Ovanesyan:2011kn}. In this paper we discuss final-state parton showers, where the virtuality and $ | \vc{k}_{\perp} | \sim Q$ are the largest in the hard scattering and it decreases with branching. This is explicitly 
taken into account in the calculation of the relevant in-medium slitting kernels that we use in this study. It is worth 
noting that in the general equation (i.e. Eq.~\eqref{gensum}) they enter as process-dependent corrections to the leading vacuum splittings. 

The full splitting functions in the medium, $P^{\rm full}_{i}$ in Eq.~\eqref{full}, govern the evolution of fragmentation functions in the medium.
Parton shower cascades in the medium behave similarly to the vacuum ones~\cite{Fickinger:2013xwa} 
and the momentum
scaling $\lambda \ll 1$ of the momentum transfers from the QCD medium~\cite{Ovanesyan:2011xy}, which are integrated over, does not change the chain of  virtuality
decay in the final state. This is actually a theoretical requirement from SCET$_{\rm G}$ when it was originally derived~\cite{Idilbi:2008vm,Ovanesyan:2011xy}. Thus,  within such a cascade picture, one might write down a generalized DGLAP-type evolution equation in the medium as follows,
\begin{eqnarray}
\frac{\d D_{h/q}(z,Q)}{\d \ln Q}&=&\frac{\alpha_s(Q)}{\pi}\int_{z}^1 \frac{\d z'}{z'}\left[P^{\rm full}_{q\rightarrow qg}(z', Q; \beta)D_{h/q}\left(\frac{z}{z'},Q\right) +P^{\rm full}_{q\rightarrow gq}(z', Q; \beta) D_{h/g}\left(\frac{z}{z'},Q\right)\right]\, ,
\label{eq:mAP10}
\\
\frac{\d D_{h/g}(z,Q)}{\d \ln Q}&=&\frac{\alpha_s(Q)}{\pi}\int_{z}^1 \frac{\d z'}{z'}\Big[P^{\rm full}_{g\rightarrow gg}(z', Q; \beta)D_{h/g}\left(\frac{z}{z'},Q\right)
+P^{\rm full}_{g\rightarrow q\bar q}(z', Q; \beta)\sum_q D_{h/q}\left(\frac{z}{z'},Q\right)\Big]\, .
\label{eq:mAP30}
\end{eqnarray}
A couple of comments are on order. First, since we follow the same cascade approximation, the modified DGLAP evolution equations in the medium have the same form as those in the vacuum in Eq.~\eqref{eq:AP10} and \eqref{eq:AP30}, with only the vacuum splitting functions $P_i^{\rm vac}$ replaced by the full splitting functions $P_i^{\rm full}$. Second, such modified evolution equations still have the logarithmic $\ln Q$-enhancement. This is because the full medium-induced splitting functions are equal to the sum of the vacuum and the medium-induced ones, as in Eq.~\eqref{full}. The vacuum ones are always accompanying $1/\vc{k}_{\perp}^2$ as in Eq.~\eqref{qqg} and thus have the logarithmic enhancement. At the same time, the medium-specific term, i.e., $\frac{1}{{\vc{k}}_{\perp}^2} P^{\text{med,~real}}_{i}(x,\vc{k}_{\perp})$ is finite  in the strict $\vc{k}_\perp\to 0$ limit due to the Landau-Pomeranchuk-Migdal effect~\cite{Ovanesyan:2011kn}. 
Eventually, since the full medium-induced splittings  are the sum of the two, they thus still contain the logarithmic enhancement from the large $Q^2$ phase space. This justifies the usage of a similar DGLAP-type evolution equation in the medium, which is true at least within the cascade approximation we are taking.  Similar studies along this line but within different 
theoretical framework has been considered in, e.g., Refs.~\cite{Wang:2001ifa,Chang:2014fba}.  In the next subsection, we will see how these evolution equations encode the effect of parton energy loss from multiple soft gluon emissions.

\subsection{From QCD evolution to energy loss}
In the soft-gluon emission limit, the approximate solutions to the evolution equations for fragmentation functions are connected to  the energy-loss of quarks and gluons. In this limit, the evolution equations in both the vacuum and the medium simplify and decouple,
\begin{equation}
\frac{\d D_{h/c}(z,Q)}{\d\ln Q}= \frac{\alpha_s}{\pi}\int_z^{1}\frac{\d z'}{z'}\,\left[P_{c\to cg}(z',Q)\right]_+D_{h/c}(z/z',Q).
\label{eq:simple}
\end{equation}
The vacuum DGLAP evolution equations can be written explicitly as
\begin{eqnarray}
\frac{\d D^{\rm vac}_{h/c}(z, Q)}{\d\ln Q}&=& 2C_R \frac{\alpha_s}{\pi} \left\{\int_z^{1}d z' \frac{1}{1-z'}\left[\frac{1}{z'}\,D^{\rm vac}_{h/c}(z/z',Q)-D^{\rm vac}_{h/c}(z, Q)\right]  \right. + D^{\rm vac}_{h/c}(z, Q) \ln(1-z)  \Big\} \, ,
\end{eqnarray}
and we expand the integrand around the end point  $z' = 1$. Note that $D^{\rm vac}_{h/c}(z,Q)$ falls steeply with increasing $z$, and we define 
\beaa
n(z) = -\frac{d\,\ln D^{\rm vac}_{h/c}(z)}{d\,\ln z}
\eeaa
as a measure of the steepness of the fragmentation functions. To modified leading logarithmic accuracy, we neglect the $Q$-dependence in $n(z)$ and use the one-loop running of the strong coupling $\alpha_s(Q) = 1/(b_0\ln\frac{Q^2}{\Lambda_{\rm QCD}^2})$ where $b_0 = (11C_A - 4n_f T_R)/(12\pi)$. The evolution equations can be solved in closed forms,
\beaa
D_{h/c}^{\rm vac}(z, Q) = \exp \left[ - \frac{C_R}{\pi}   \frac{1}{b_0}  \ln \left( \frac{\ln  \frac{Q}{\Lambda_{\rm QCD}} }{ \ln \frac{Q_0}{\Lambda_{\rm QCD}} }
\right)   \left\{[n(z)-1](1-z) - \ln (1-z) \right\} \right] D^{\rm vac}_{h/c}(z, Q_0).
\label{eq:sol-vac}
\eeaa

Following a similar procedure, we find that the solutions to the generalized DGLAP  evolution equations in the medium are,
\beaa
D_{h/c}^{\rm med}(z, Q) =&
\exp \left[ - \frac{C_R}{\pi}   \frac{1}{b_0}  \ln \left( \frac{\ln  \frac{Q}{\Lambda_{\rm QCD}} }{ \ln \frac{Q_0}{\Lambda_{\rm QCD}} }
\right)   \left\{[n(z)-1](1-z) - \ln (1-z) \right\} \right] D^{\rm vac}_{h/c}(z, Q_0)
\nonumber\\
&\times \exp\left[-[n(z) - 1] \left\{\int_0^{1-z} dz' z'\, \int_{Q_0}^Q dQ' \frac{dN}{dz'dQ'}\right\} - \int_{1-z}^1 dz' \int_{Q_0}^Q \frac{dN}{dz'dQ'}\right].
\label{eq:med-sol}
\eeaa
This equation can be further written as
\beaa
D_{h/c}^{\rm med}(z, Q) = D_{h/c}^{\rm vac}(z, Q)\, \exp\left[ -[n(z)-1] \left\langle \frac{\Delta E}{E}\right \rangle_z - \left \langle N_g\right \rangle_z\right],
\label{eq:sol-med}
\eeaa
where 
\beaa
\left\langle\frac{\Delta E}{E}\right\rangle_z&=\int_0^{1-z}\d x \,x\,\frac{\d N}{\d x}(x)  ,
\\
\left\langle N_g\right\rangle_z&=\int_{1-z}^1\d x\,\frac{\d N}{\d x}(x).
\eeaa
Note that the vacuum evolution and medium correction effects factorize, as can be seen in Eqs.~\eqref{eq:med-sol} and \eqref{eq:sol-med}, and the medium causes an attenuation of the fragmentation functions. Also, both $\left\langle\frac{\Delta E}{E}\right\rangle_z$ and $\left\langle N_g \right\rangle_z$ depend on the scale $Q$ through,
\beaa
\frac{dN}{dx} = \int_{Q_0}^Q dQ'\,\frac{dN}{dx'dQ'},
\eeaa
and
\begin{equation}
\left\langle\frac{\Delta E}{E}\right\rangle_z \xrightarrow{z \to 0}\left\langle\frac{\Delta E}{E}\right\rangle\, ,
\qquad
\left\langle N_g\right\rangle_z \xrightarrow{z\to1}\left\langle N_g\right\rangle\,.
\end{equation}
For small to intermediate values of $z$, the suppression of the fragmentation functions (and hence the inclusive hadron suppression factor) is controlled by the fractional energy loss $\left\langle\frac{\Delta E}{E}\right\rangle$ and the steepness $n(z)$ of fragmentation functions. On the other hand, for values of $z\approx 1$, the suppression factor becomes $\sim \exp(-\langle N_g \rangle )$ which is the probability of not emitting a gluon. Both behaviors are the same as in the energy loss formalism. This shows the connection between the QCD evolution approach and the energy loss formalism in the small-$x$ approximation.


\section{Inclusive hadron suppression in  Pb+Pb collisions  at the LHC}
\label{sec:hadron}

In this section we compare our calculations to the LHC Pb+Pb run I data (A+A), and make predictions for run II (A+A). We focus on single inclusive hadron production in Pb+Pb collisions of the form $A+B\to h(p_T, y)+X$, where $A$ and $B$ are the incoming nuclei, and $h$ is the hadron with transverse momentum $p_T$ and rapidity $y$, respectively. In the first part of this section, we compare with the experimental data in Pb+Pb run I at a center-of-mass energy per nucleon-nucleon pair $\sqrt{s_{NN}}=2.76$ TeV. The second part is devoted to the theoretical predictions for the Pb+Pb run II at $\sqrt{s_{NN}}=5.1$ TeV. The nuclear modification of single inclusive hadrons in Pb+Pb collisions can be quantified through the nuclear modification factor $R_{AA}$:
\beaa
R_{AA} (p_T)  =  \frac{ d \sigma^h_{AA}/dyd^2 p_T }{\langle N_{ \rm coll}\rangle d \sigma^h _{pp}  / dyd^2 p_T},
\eeaa
where $\langle N_{\rm coll}\rangle$ is the average number of binary nucleon-nucleon collisions and is computed at a given centrality, and $d \sigma^h_{AA}/dyd^2 p_T$ and $d \sigma^h _{pp}  / dyd^2 p_T$ are the differential cross section of inclusive hadron production in A+A and p+p collisions, respectively. Unless otherwise specified, we focus on the single hadron production around the mid-rapidity region, and consider two centrality classes: (1) the most central collisions (0-10\%) with average number of participants $N_{\rm part} = 350$; (2) the mid-peripheral collisions (30-50\%) with average number of participants $N_{\rm part} = 110$. 
Experimental results are often presented in different centrality classes and, for comparison, we show data sets with $N_{\rm part}$  closest to our calculations.

The invariant inclusive hadron production cross section in p+p collisions can be written as
\beaa
\frac{d \sigma^h _{pp}}{dyd^2 p_T} = \sum_{c} \int \frac{dz}{z^2}
\frac{d \hat \sigma_c \left(p_{T_c}=p_T/z\right)}{dyd^2 p_{T_c}} D_{h/c}^{\rm vac}(z, Q), 
\label{convpp}
\eeaa
while the corresponding cross section in A+A collisions reads
\beaa
\frac{1}{\langle N_{\rm coll}\rangle} \frac{d \sigma^h _{AA}}{dyd^2 p_T} = \sum_{c} \int \frac{dz}{z^2}
\frac{d \hat \sigma_c^{\rm CNM} \left(p_{T_c}=p_T/z\right)}{dyd^2 p_{T_c}} D_{h/c}^{\rm med}(z, Q).
\label{convaa}
\eeaa
In both equations, $c=\{q, \bar q, g\}$, and $D_{h/c}^{\rm vac}(z, Q)$ and $D_{h/c}^{\rm med}(z, Q)$ are the parton-to-hadron fragmentation functions in vacuum and QGP medium evolved between the 
hard and lowest  scales, respectively. They are obtained from the vacuum and medium evolution equations, as given in Eqs.~(\ref{eq:AP10}-\ref{eq:AP30}) and (\ref{eq:mAP10}-\ref{eq:mAP30}), respectively. For the vacuum input, we take KKP fragmentation functions~\cite{Kniehl:2000fe}, and choose the factorization, fragmentation, and renormalization scales $Q=p_{T_c}$ in the following numerical calculations.

\begin{figure}[!t]
\centering
\includegraphics[width=0.47\textwidth]{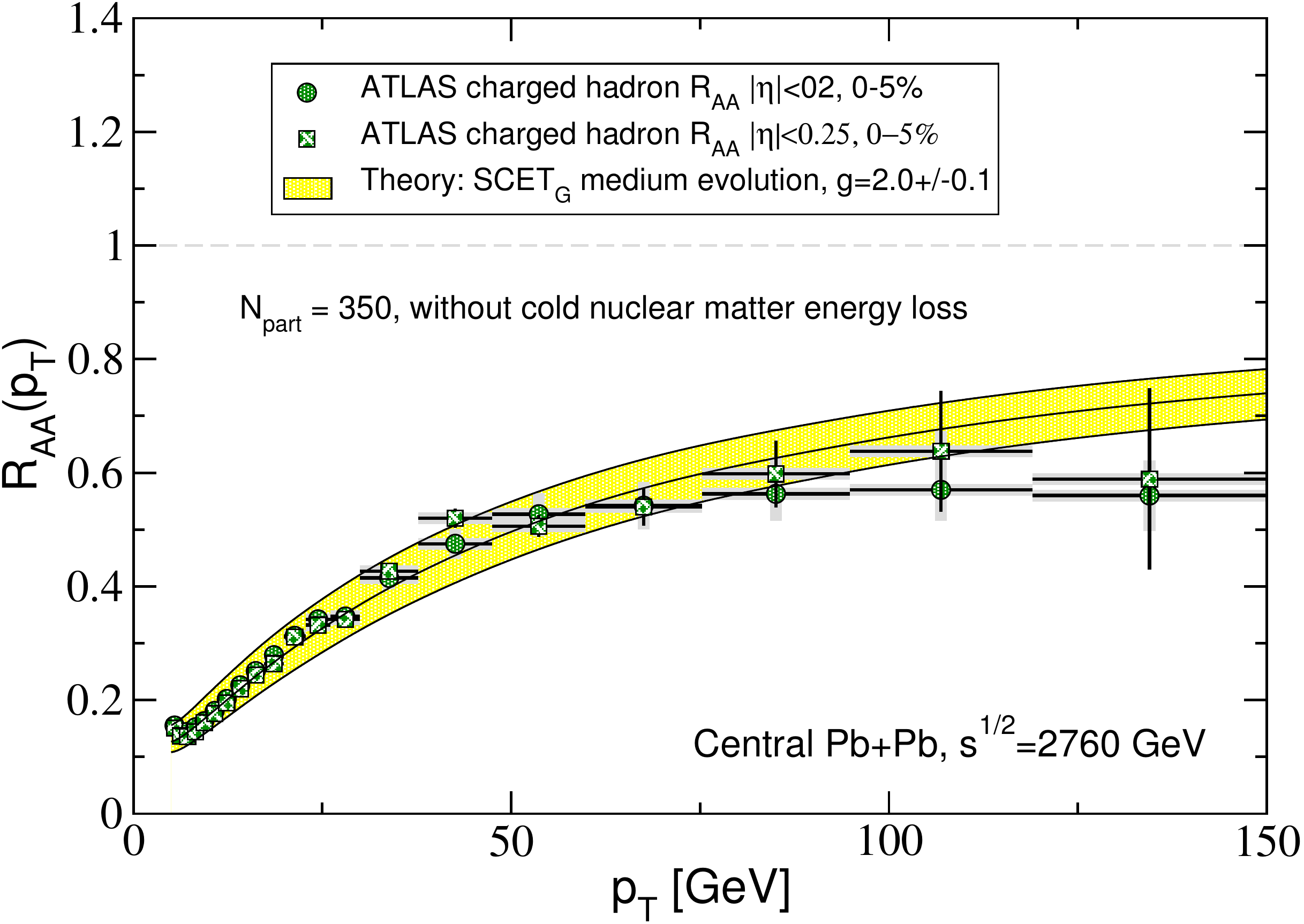}
\hskip 0.2in
\includegraphics[width=0.47\textwidth]{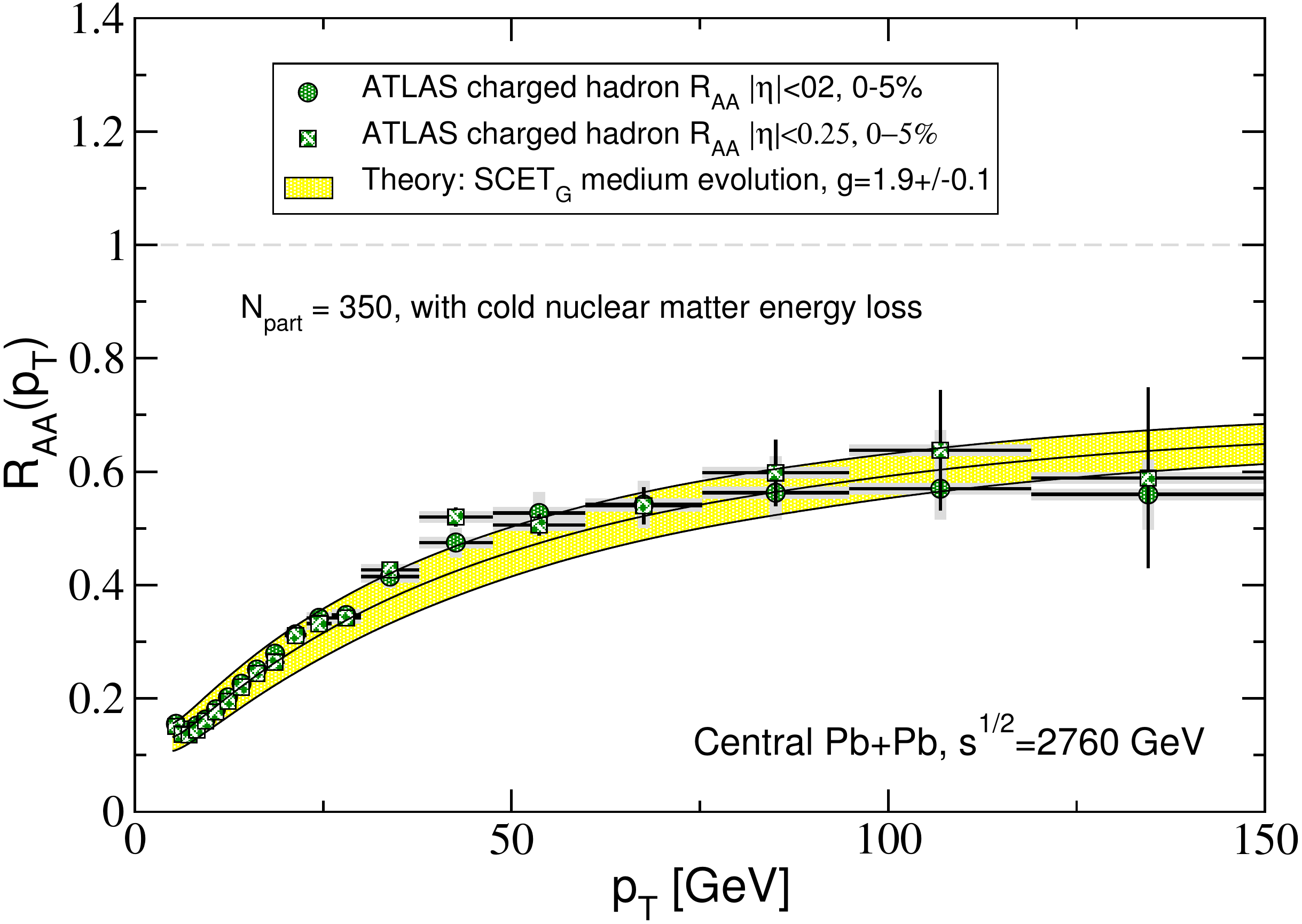}\\
\includegraphics[width=0.47\textwidth]{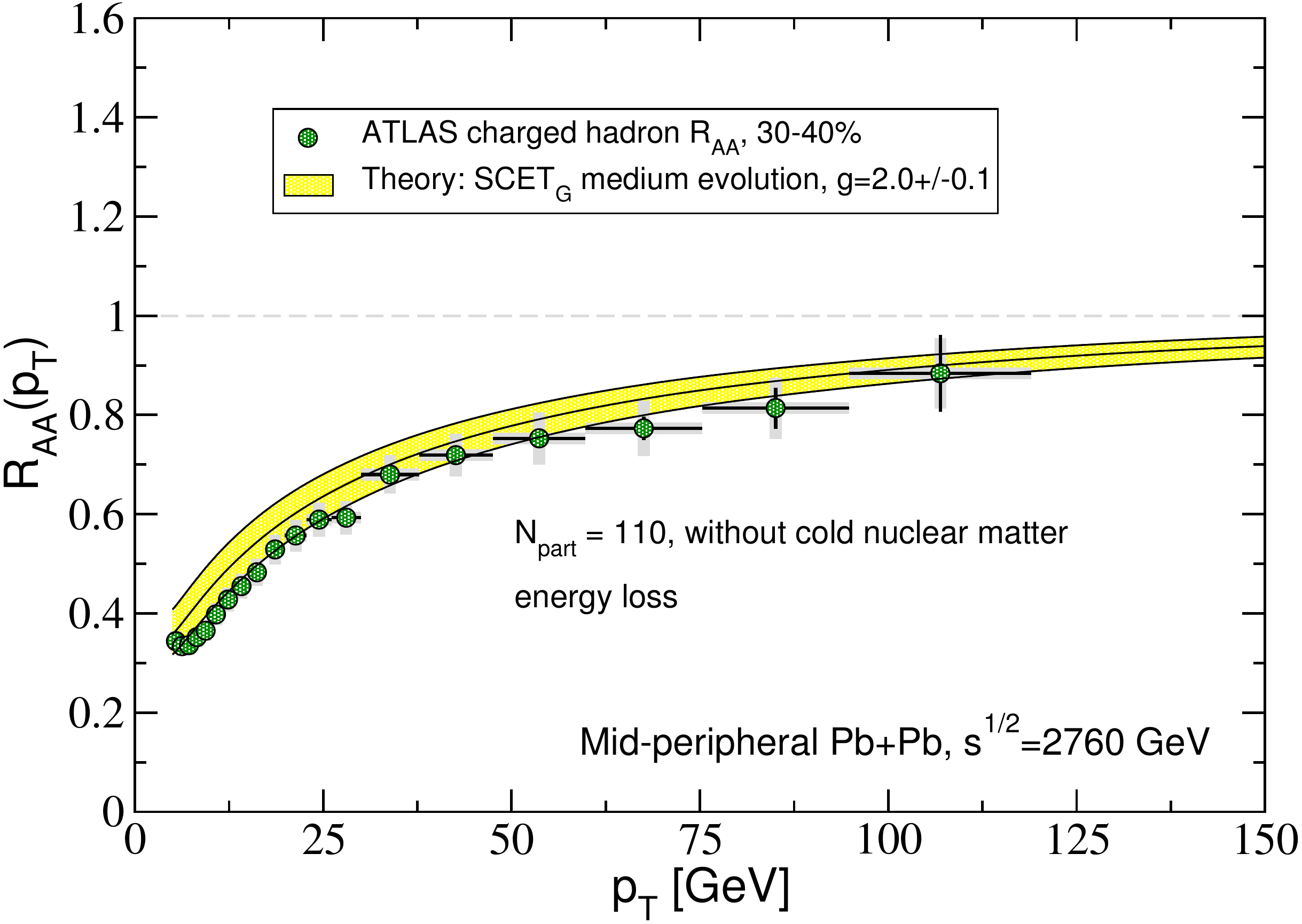}
\hskip 0.2in
\includegraphics[width=0.47\textwidth]{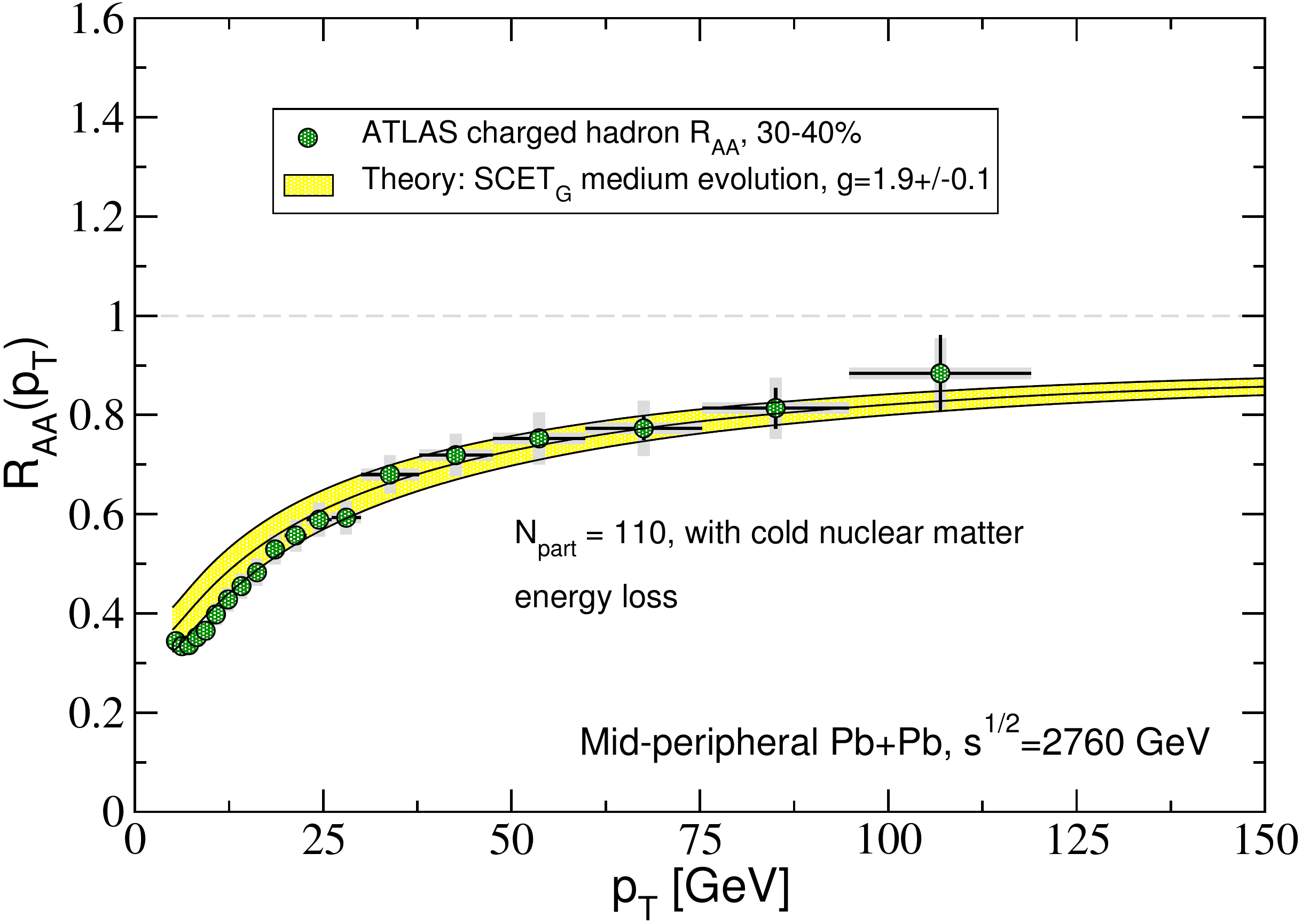}\\
\includegraphics[width=0.47\textwidth]{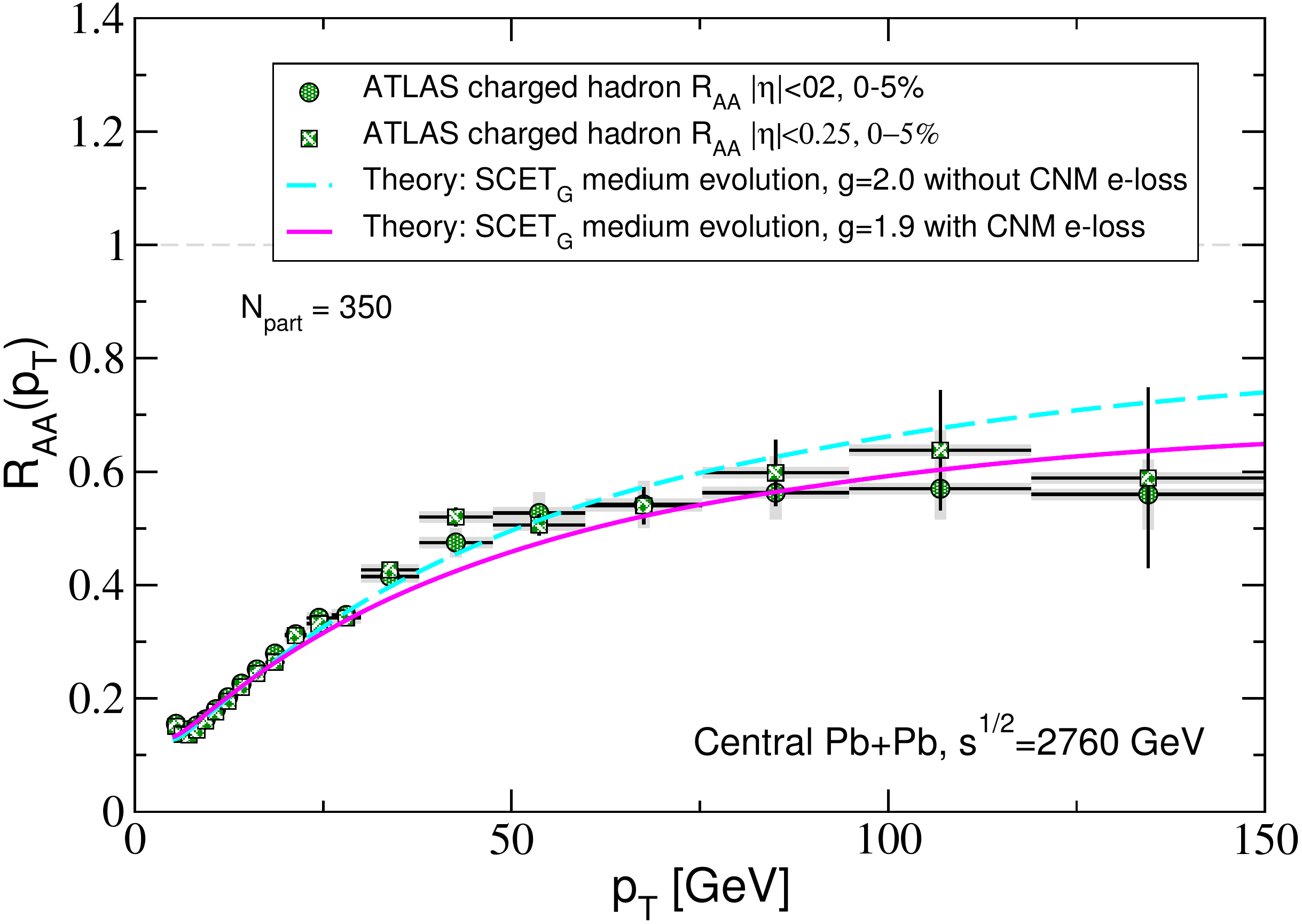}
\hskip 0.2in
\includegraphics[width=0.47\textwidth]{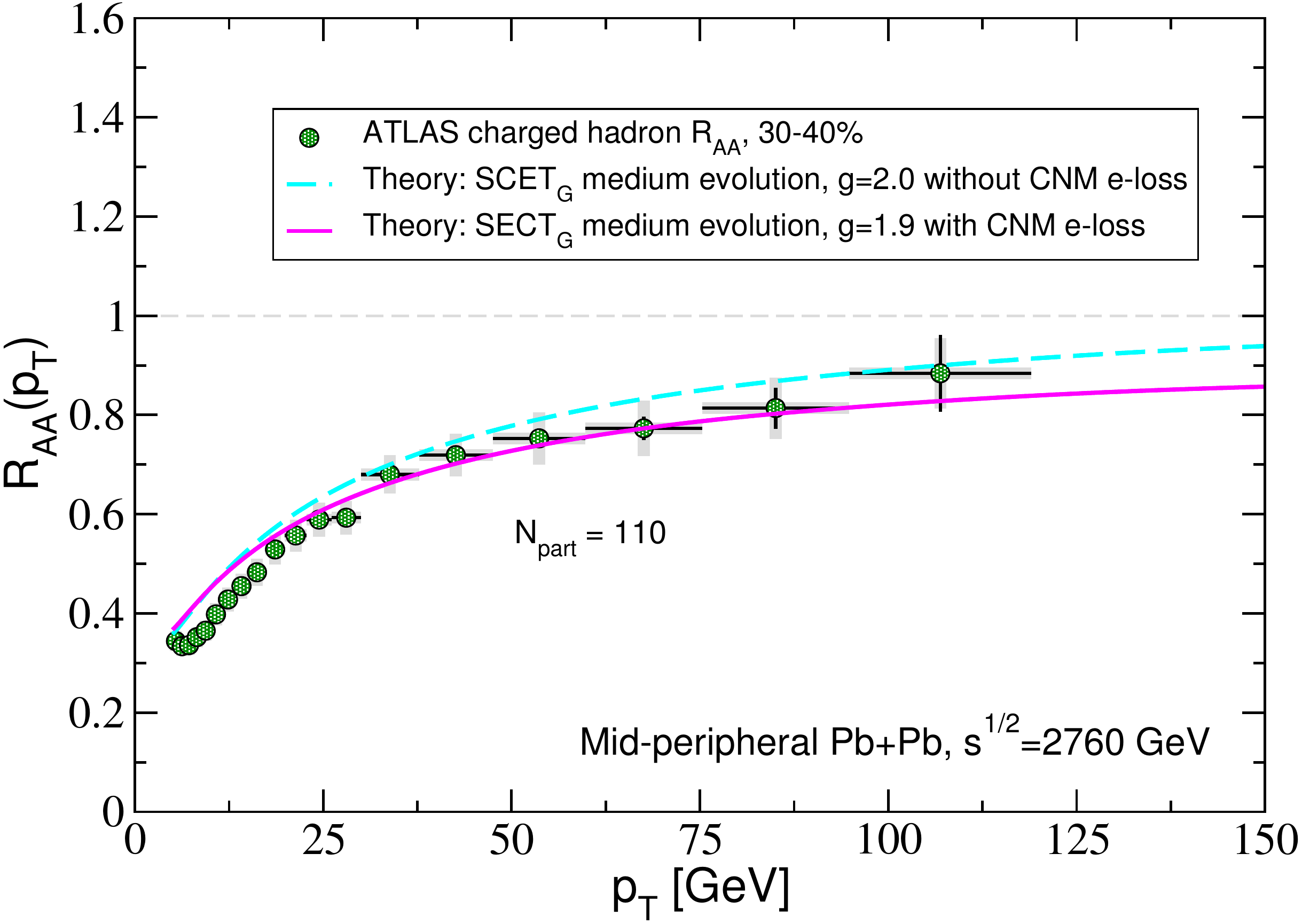}
\caption{The nuclear modification factor $R_{AA}$ for charged hadrons is calculated in central Pb+Pb collisions at the LHC $\sqrt{s_{NN}}=2.76$ TeV, and compared with the ATLAS experimental data~\cite{Aad:2015wga}. Top left panel: central collisions without CNM energy loss, and bands represent the variation of the coupling strength $g=2.0\pm 0.1$. Top right panel: central collisions with CNM energy loss, and bands represent the variation of the coupling strength $g=1.9\pm 0.1$. Middle panels are the same as the top panels but calculated in mid-peripheral Pb+Pb collisions. Bottom panels show the comparison of our calculations with ATLAS charged hadron suppression measurements in central (left), as well as mid-peripheral (right) Pb+Pb collisions. The dashed cyan curves are for the coupling strength $g=2.0$ without CNM energy loss, while the solid magenta curves are for the coupling strength $g=1.9$ with CNM energy loss.}
\label{ATLAS-eloss}
\end{figure}

In Eq.~\eqref{convaa} $d\sigma_c \left(p_c=p_T/z\right)/dyd^2 p_{T_c}$ is the hard parton production cross section, while $d\sigma_c^{\rm CNM} \left(p_c=p_T/z\right)/dyd^2 p_{T_c}$ is the same hard parton production cross sections calculated with isospin and initial-state cold nuclear matter (CNM) effects.  All CNM effects considered, which include  dynamical nuclear shadowing~\cite{Qiu:2004qk}, Cronin effect~\cite{Accardi:2002ik,Kang:2014hha} and initial-state parton 
energy loss~\cite{Gavin:1991qk,Vitev:2007ve,Xing:2011fb}, have clear physical/dynamical origin  centered around the picture of multiple parton scattering~\cite{Vitev:2006bi}, see Appendix~D for implementation.

In p+A collisions there is one target nucleus to generate CNM effects through the elastic, inelastic, and coherent
scattering of the incoming and outgoing partons. In A+A there are two nuclei and we expect that the cross 
section modification will be a superposition of the corresponding modifications. This does not necessarily imply
that the cross section modification in A+A reactions due to CNM effects is simply twice the  cross section 
modification in p+A reactions. The precise numerical value depends on the rapidity $y$ of the produced particles. For example, at very forward rapidity 
the Bjorken momentum fraction of one of the hard scattering partons is 
$x_a \sim 1$, which amplifies the effect of cold nuclear nuclear matter energy loss but eliminates coherent power 
corrections. The opposite is true for the second parton with Bjorken momentum fraction  $x_b \ll 1$. 
It is only at midrapidity, due to the  symmetry in the problem, that the nuclear modification due to cold 
nuclear matter effects in A+A  is roughly double the nuclear modification expected in p+A collisions~\cite{Vogt:2015uba,Vitev:2002pf}. 
An alternative approach would be to include CNM effects 
in nuclear parton distributions, see e.g. ~\cite{Kovarik:2015cma,Gauld:2015lxa}.

In the region of interest  
$p_T > 5$~GeV    dynamical nuclear shadowing/power corrections are negligible. Cronin effect can play a  role up to 
$p_T \simeq 10$~GeV. The effect of initial-state  cold nuclear matter (CNM) energy loss can be amplified near kinematic 
bounds~\cite{Neufeld:2010dz} and is the only cold nuclear matter effect relevant to our high $p_T$ phenomenology.
 While at moderate transverse momenta 
ALICE measurements in $\sqrt{s_{NN}}=5.02$~TeV p+Pb collisions~\cite{ALICE:2012mj}  are only compatible with small cold nuclear matter energy loss~\cite{Kang:2012kc}, 
ATLAS jet measurements~\cite{ATLAS:2014cpa} for the same system allow for considerably larger CNM effects~\cite{Kang:2015mta}. 
It is, thus, important to explore at least a limited  range of possibilities for cold nuclear matter energy loss.

\subsection{Comparison to LHC Pb+Pb run I data }



We are now ready to compare our calculations to the existing experimental data from the LHC Pb+Pb run I on single inclusive hadron production.
We will present calculations both with and without cold nuclear matter energy loss. There is a trade off  in the
nuclear modification factor $R_{AA}$  of  inclusive hadrons between the in-medium modification of fragmentation
functions, driven by the coupling $g$ between the hard partons and the QGP medium,  and the strength of cold nuclear matter energy loss.  
In the region $p_T < 50$~GeV, where the experimental error bars are the smallest, 
a calculation with cold nuclear energy loss and smaller coupling $g$ gives very similar description to the calculation without cold nuclear matter energy loss
and slightly larger coupling $g$. We cannot favor one scenario versus the other based upon existing inclusive hadron suppression data alone.
As we will show in the next sub-section, by going to higher $p_T$ with  LHC Pb+Pb run II the differences become more pronounced and this might
provide an opportunity to better separate those scenarios. 
Of course, keeping the coupling constant $g$ fixed, for example $g=1.9$, and performing calculations with and without cold nuclear matter energy loss
 will result in slightly different magnitude of  $R_{AA}$, as we will demonstrate at the end of this subsection.

In Fig.~\ref{ATLAS-eloss}, we present our results for the nuclear modification factor $R_{AA}$ for charged hadrons compared to data from Pb+Pb collisions at the LHC $\sqrt{s_{NN}}=2.76$ TeV measured by the ATLAS collaboration. In the top row of panels we show the comparison with central (0-5\%) collisions, while the middle row shows mid-peripheral (30-40\%) collisions. For these two rows, the left panel does not include CNM energy loss, while the right panel does. The uncertainty bands represent the variation in the coupling strength between the jet and the medium. For the case without CNM energy loss it is $g=2.0\pm 0.1$, with the upper (lower) edge corresponding to $g=1.9$ ($g=2.1$).   For the case with CNM energy loss it is $g=1.9\pm 0.1$, with the upper (lower) edge corresponding to $g=1.8$ ($g=2.0$).  

As one can see, our calculations well reproduce the suppression of inclusive charged hadron production in 0-5\% central Pb+Pb collisions at the LHC measured by ATLAS collaboration. Below $p_T = 50$~GeV both evaluations give comparable description of the experimental data,  though the inclusion of the CNM energy loss leads to slightly larger curvature and, thus, better agreement with the data in the high transverse momentum  region. If other components of the calculation have negligible uncertainties, the coupling $g$ between the jet and the medium can be constrained with an accuracy of $5\%$ and the transport properties of the medium can, thus, be extracted with $20\%$ uncertainty as they scale as $g^4$. 
When comparing to the 30-40\% Pb+Pb collisions, as measured by the ATLAS collaboration, both sets of results
describe data equally well within the statistical and systematic uncertainties.

\begin{figure}[t!]
\centering
\includegraphics[width=0.47\textwidth]{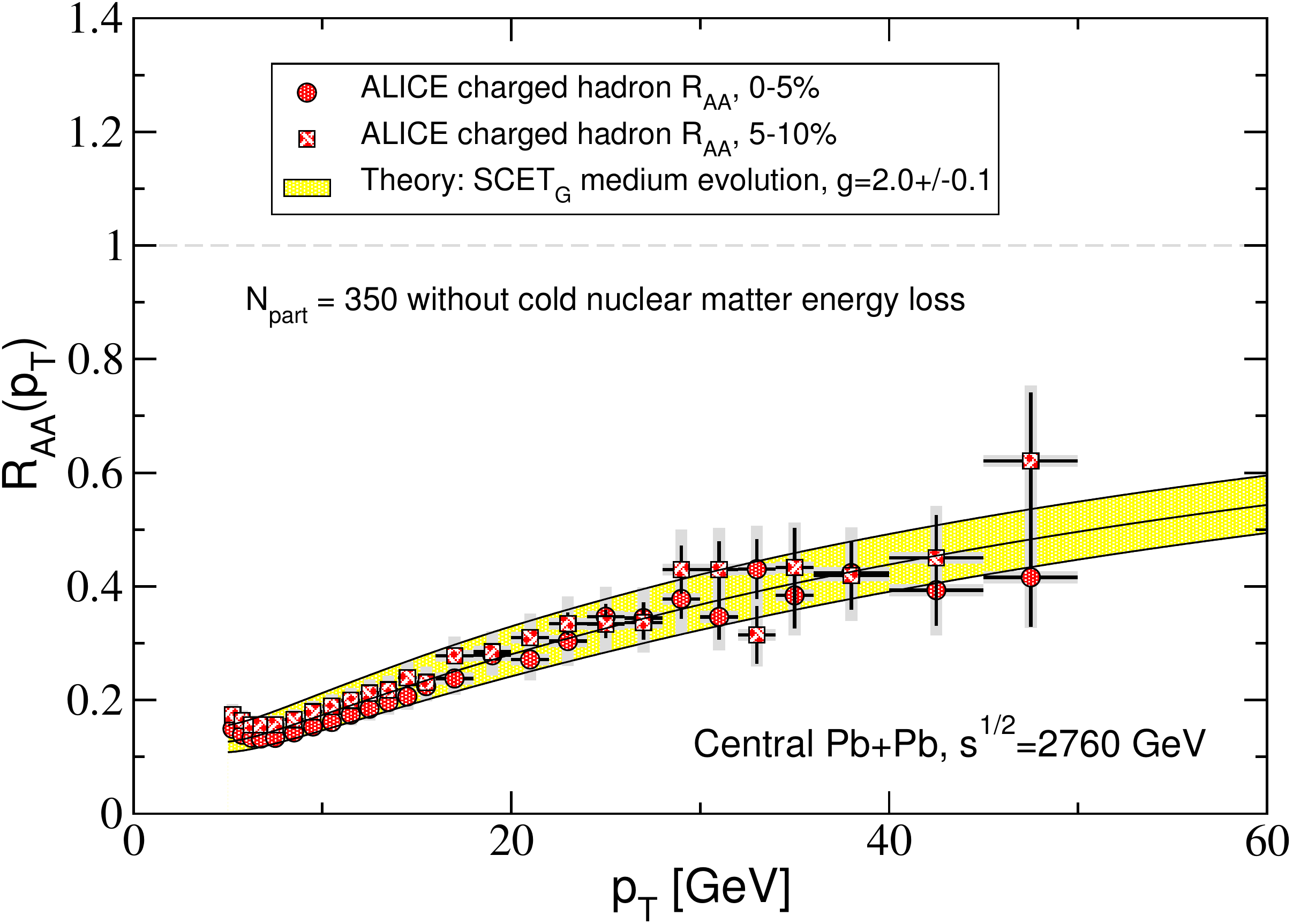}
\hskip 0.2in
\includegraphics[width=0.47\textwidth]{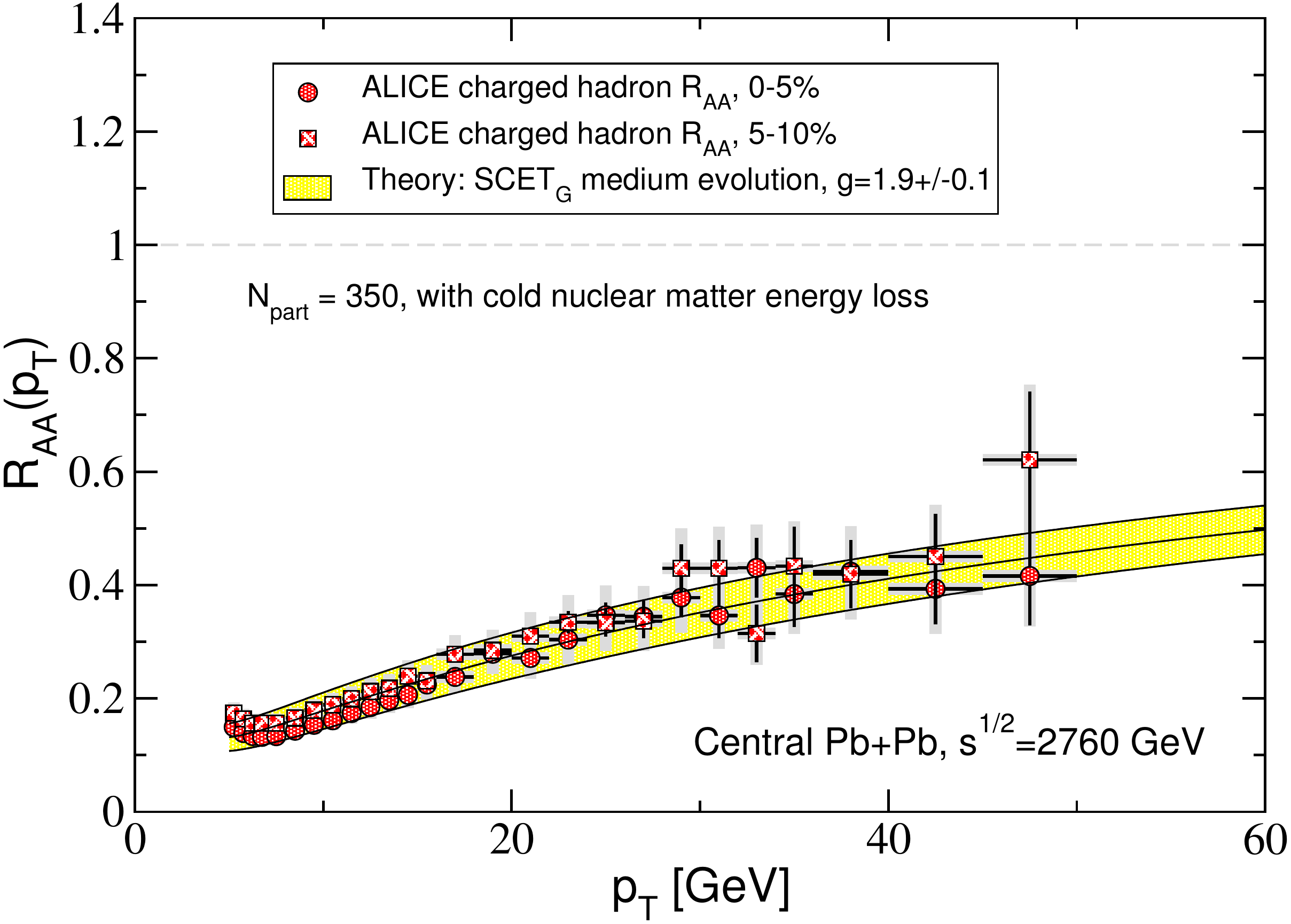}\\
\includegraphics[width=0.47\textwidth]{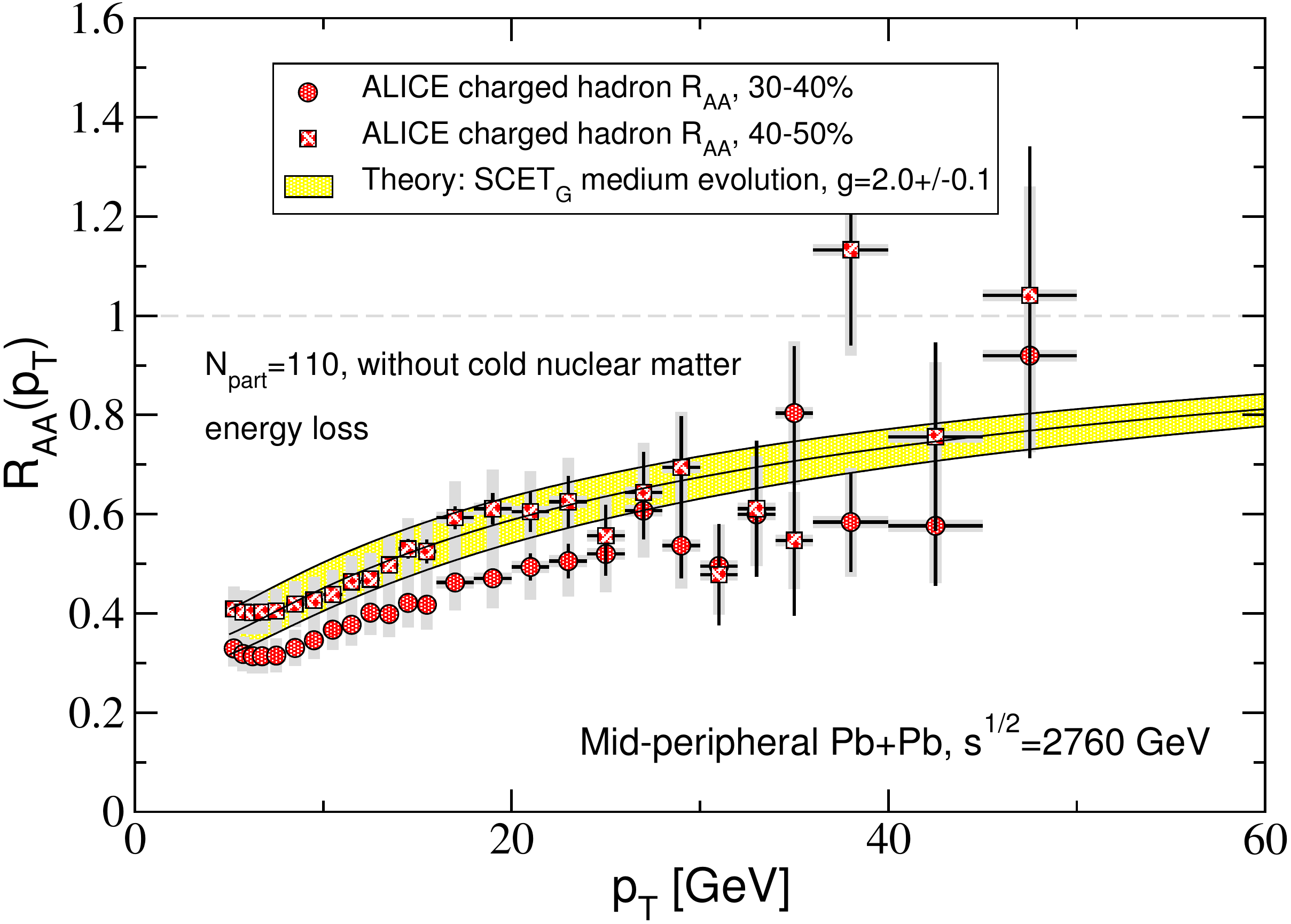}
\hskip 0.2in
\includegraphics[width=0.47\textwidth]{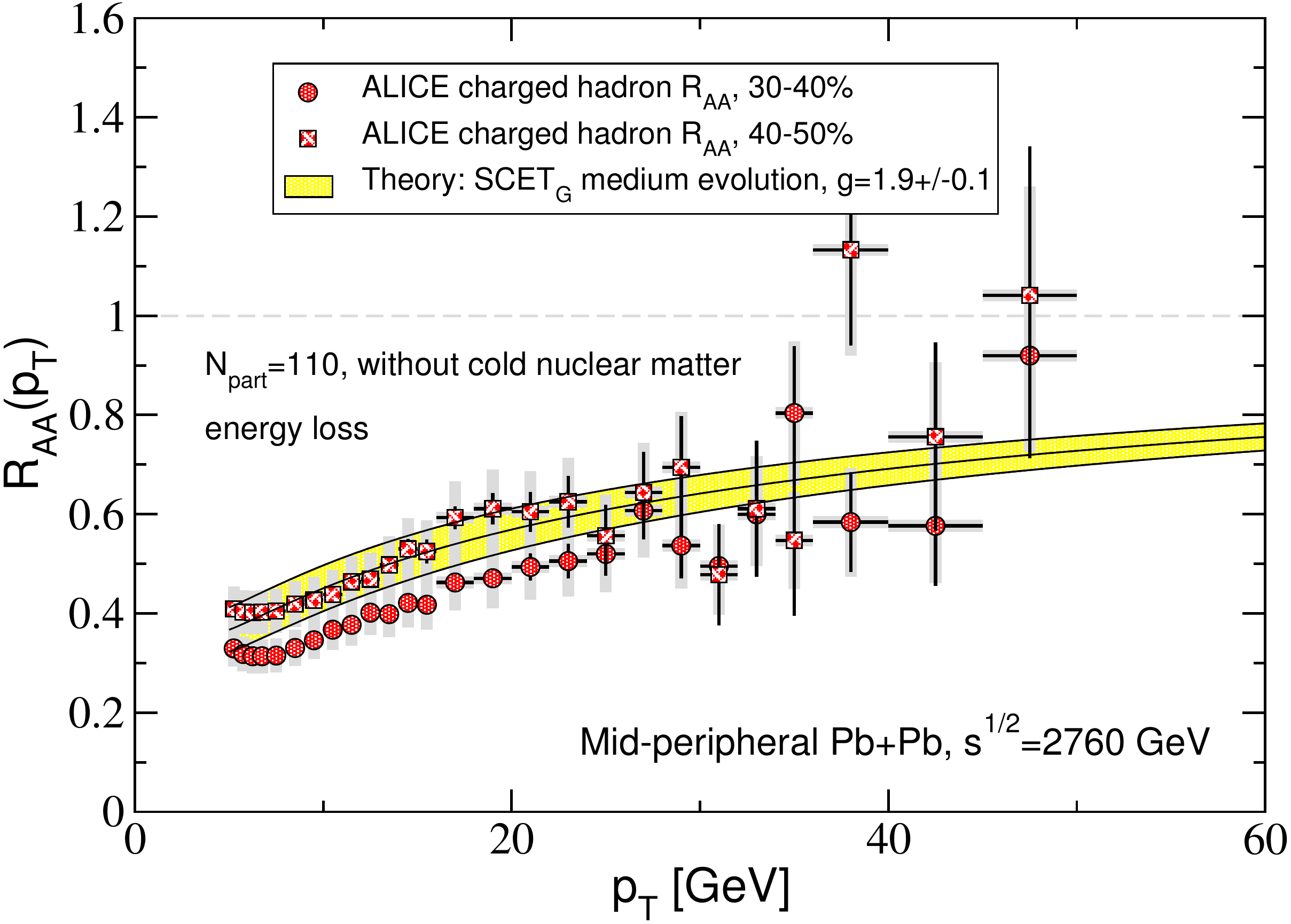}\\
\includegraphics[width=0.47\textwidth]{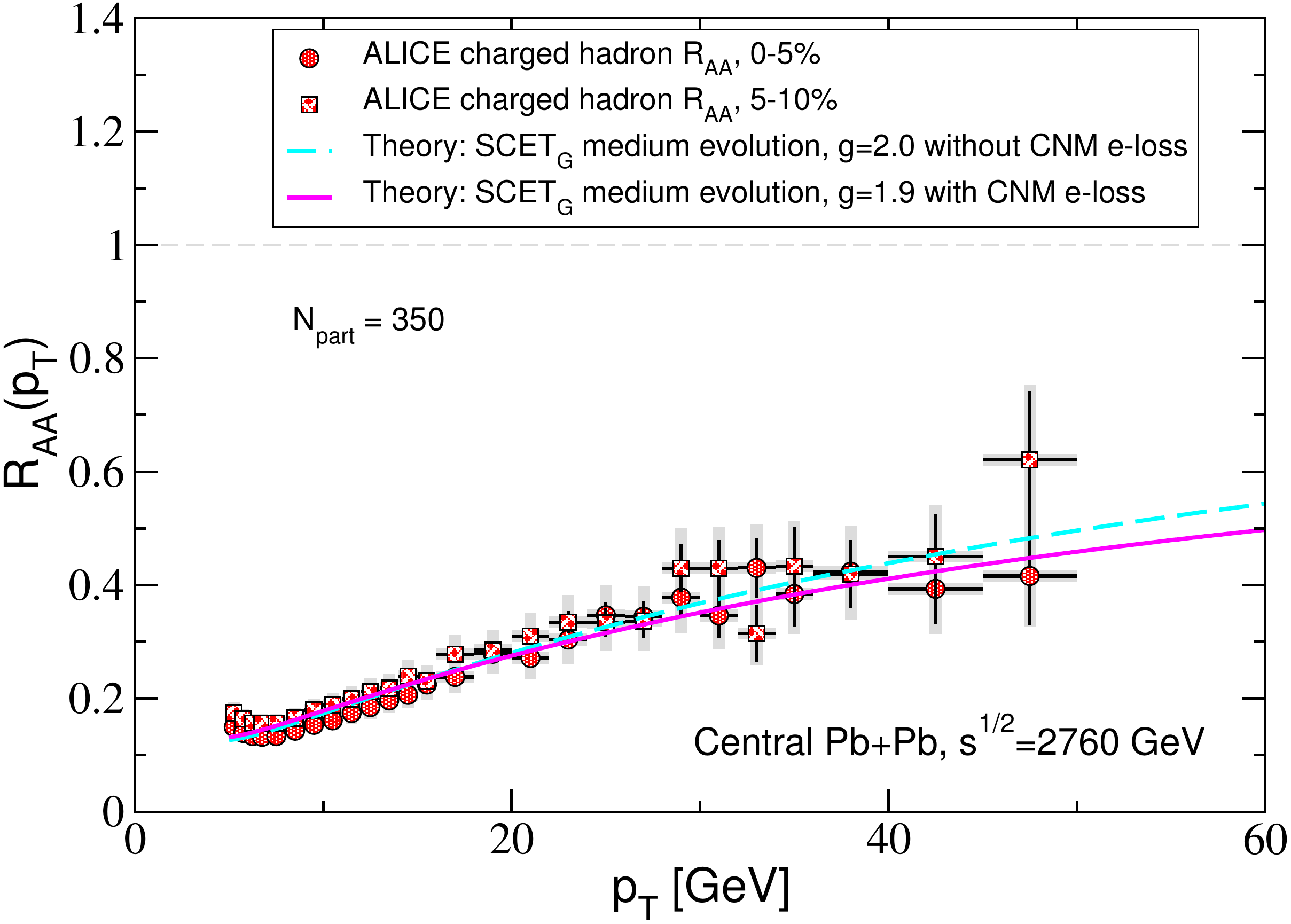}
\hskip 0.2in
\includegraphics[width=0.47\textwidth]{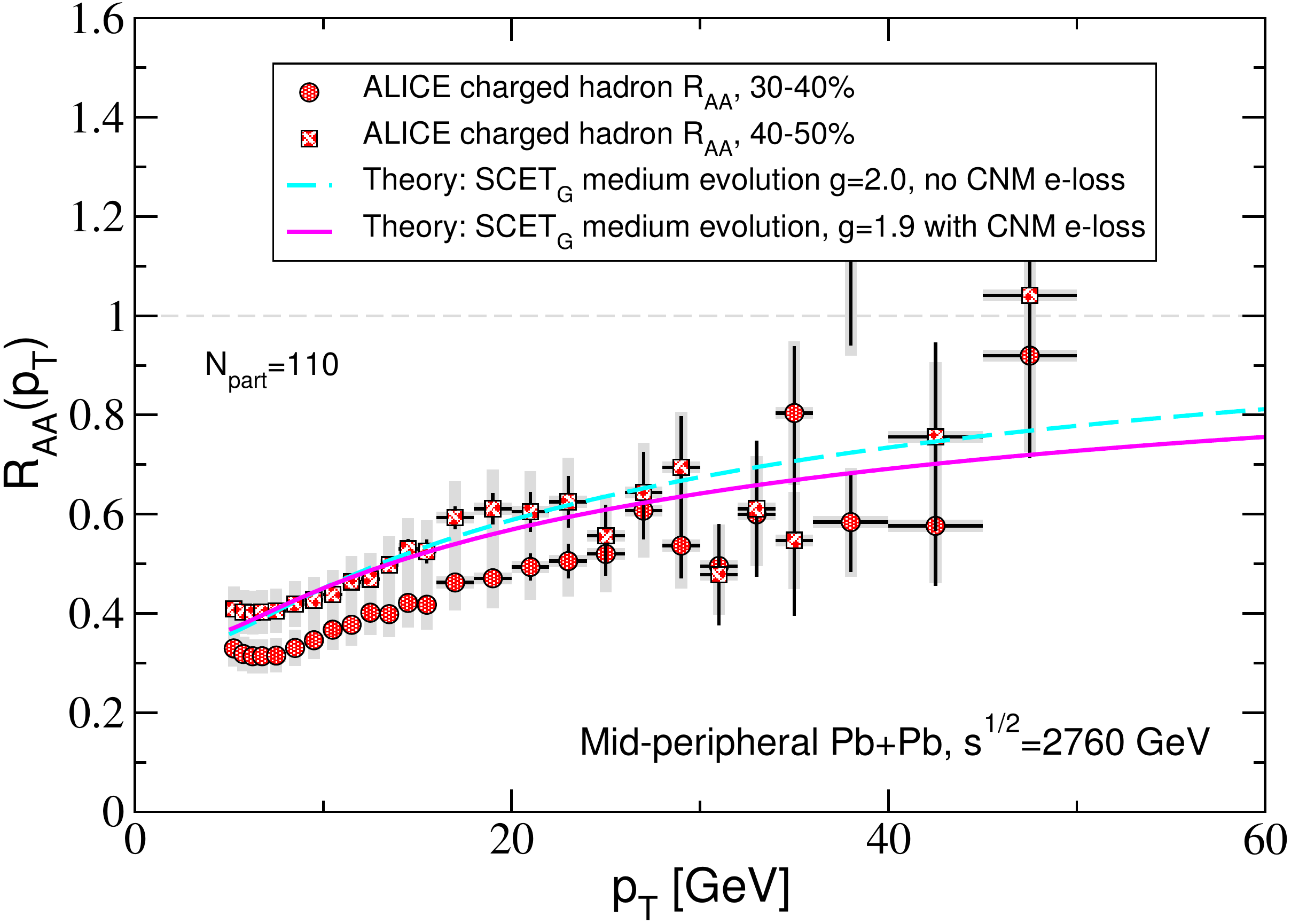}
\caption{Same as in Fig.~\ref{ATLAS-eloss} except the comparison is made to ALICE charged hadron data for the $R_{AA}$ for Pb+Pb collisions at the LHC $\sqrt{s_{NN}}=2.76$ TeV.}
\label{ALICE-eloss}
\end{figure}


\begin{figure}[!t]
\centering
\includegraphics[width=0.47\textwidth]{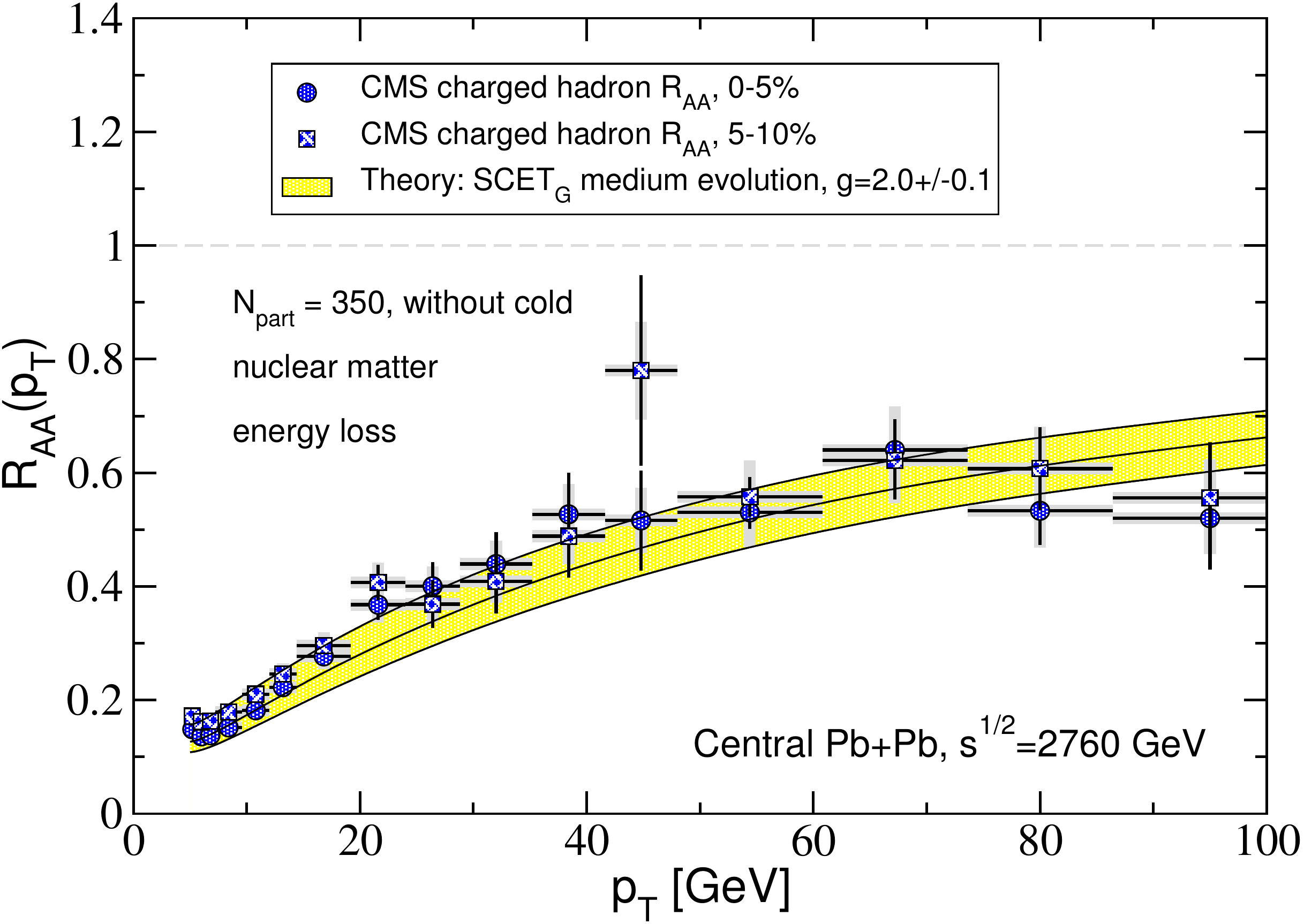}
\hskip 0.2in
\includegraphics[width=0.47\textwidth]{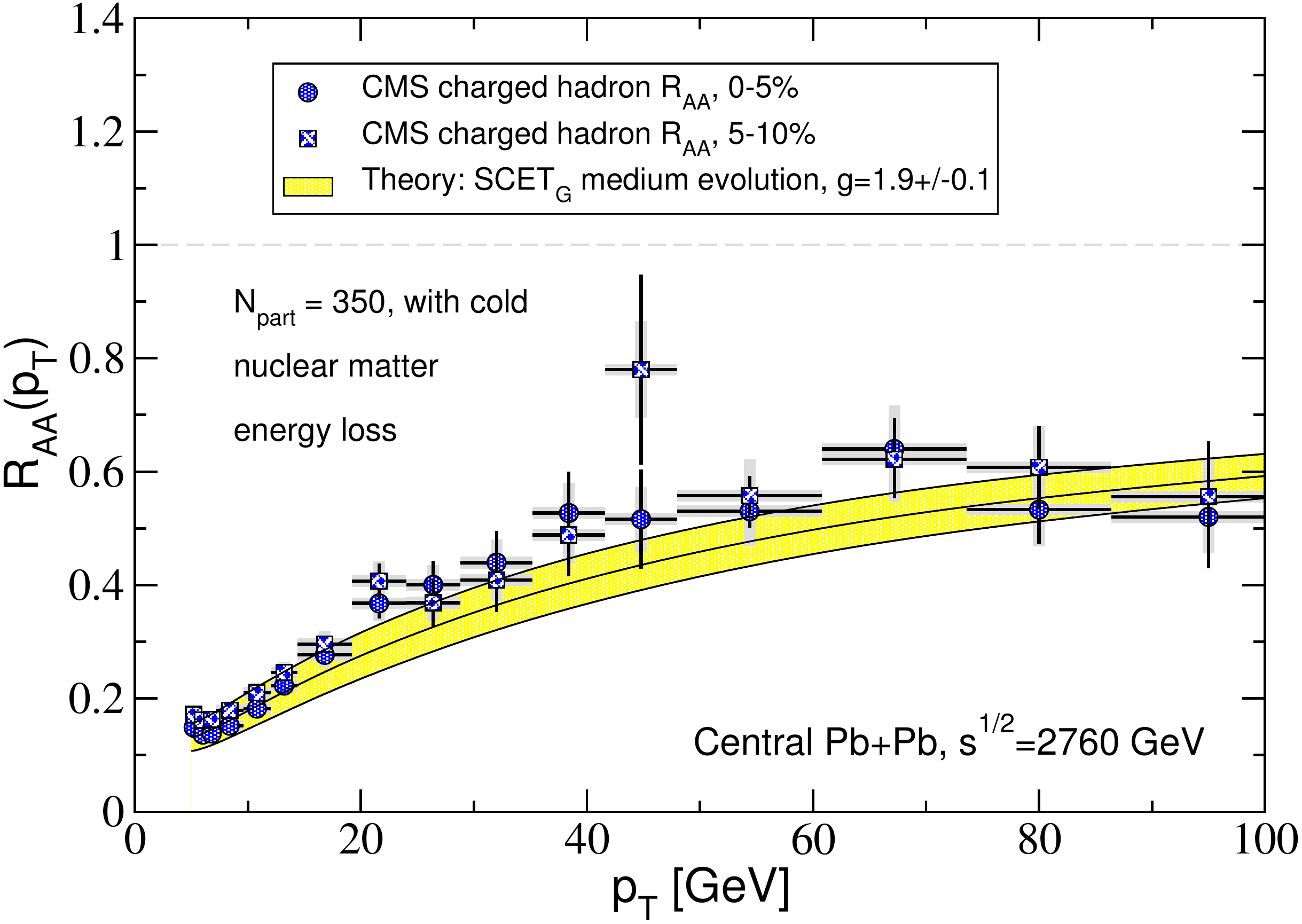}\\
\includegraphics[width=0.47\textwidth]{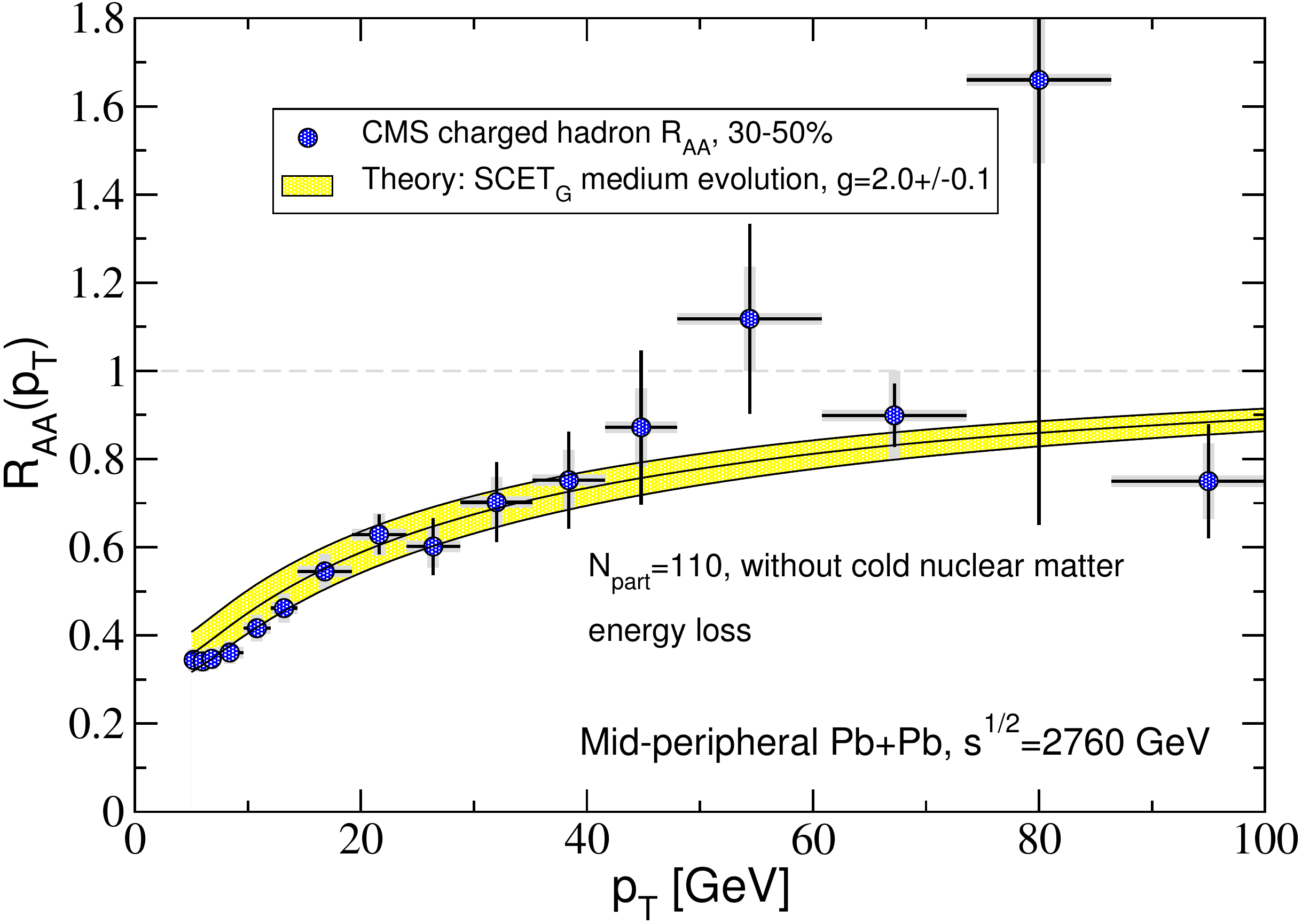}
\hskip 0.2in
\includegraphics[width=0.47\textwidth]{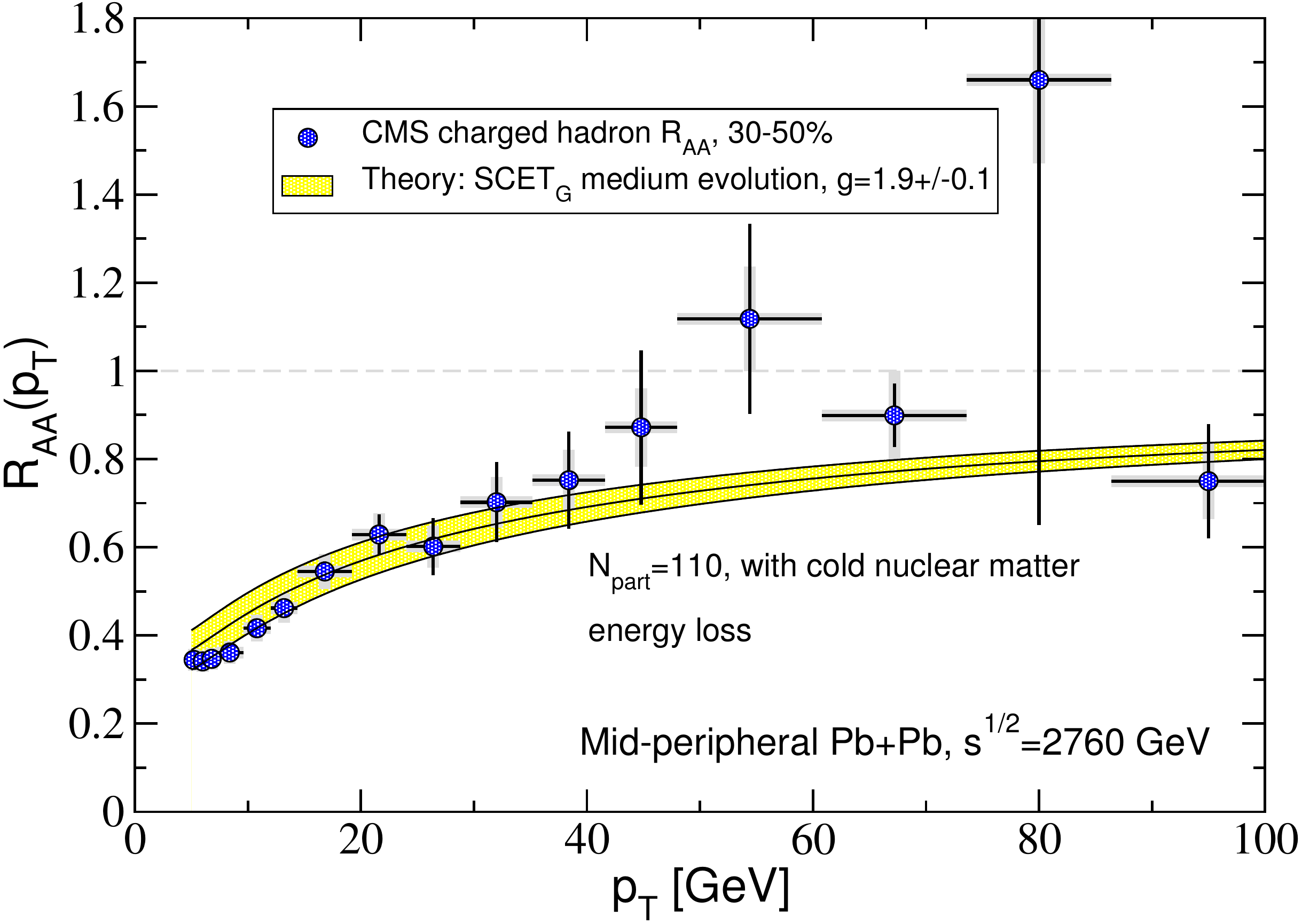}\\
\includegraphics[width=0.47\textwidth]{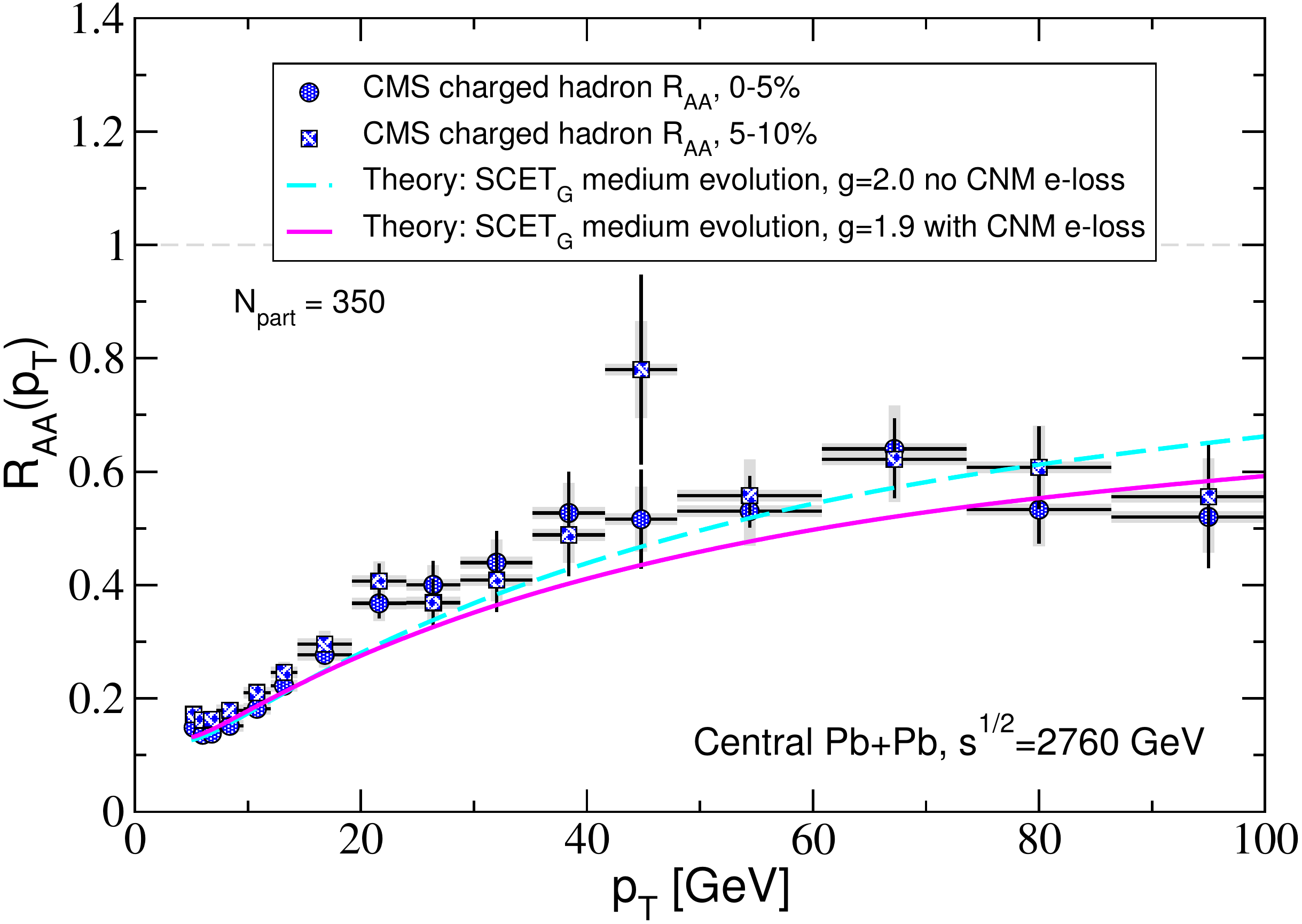}
\hskip 0.2in
\includegraphics[width=0.47\textwidth]{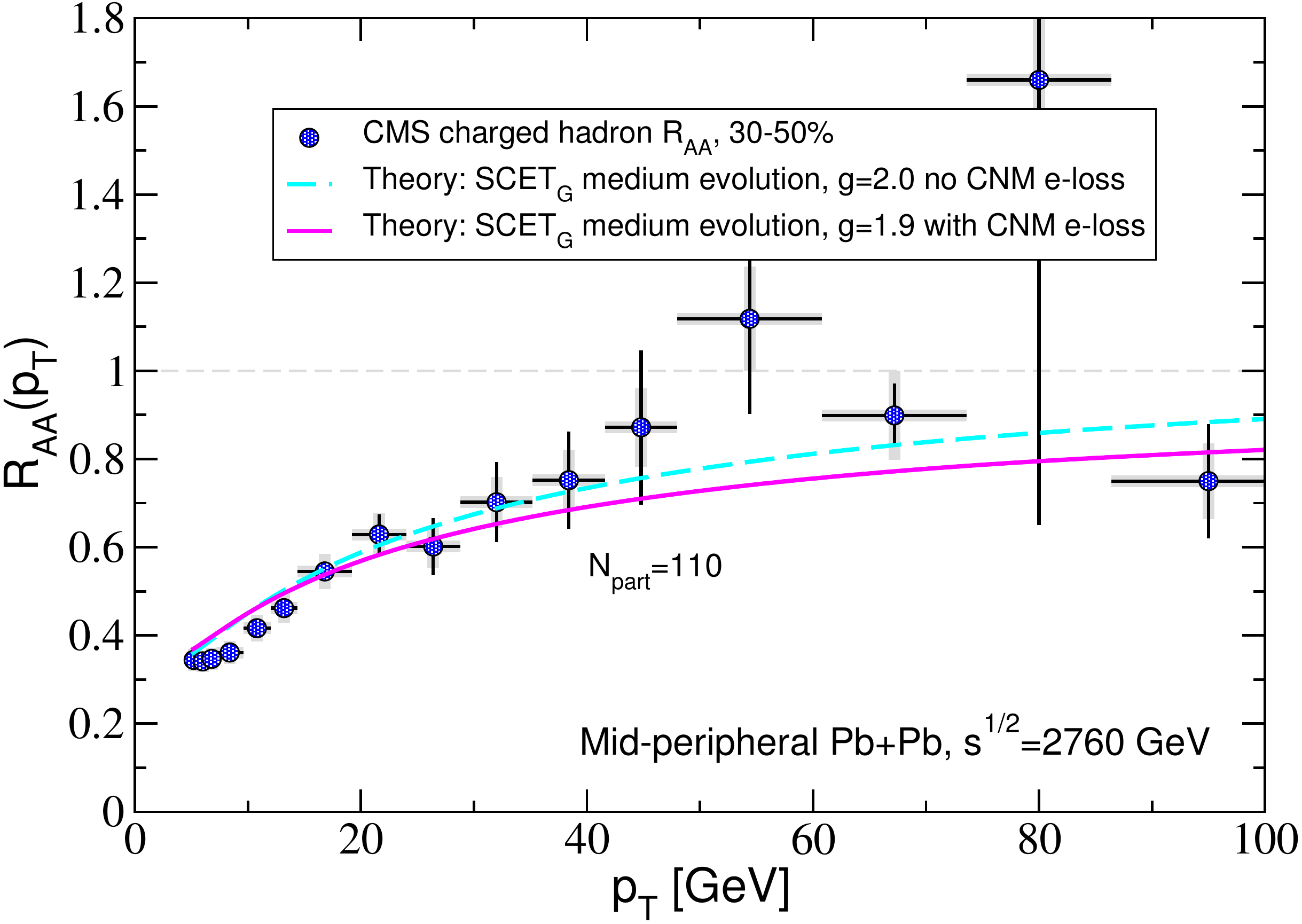}
\caption{Same as in Fig.~\ref{ATLAS-eloss} except the comparison is made to CMS charged hadron data for the $R_{AA}$ for Pb+Pb collisions at the LHC $\sqrt{s_{NN}}=2.76$ TeV.
}
\label{CMS-eloss}
\end{figure}

To better understand the relative contribution of initial-state CNM energy loss and final-state QGP effects, we combine the calculations with different combinations of $g$ and CNM energy loss  in the bottom two panels of Fig.~\ref{ATLAS-eloss}, for both central (left) and mid-peripheral (right) Pb+Pb collisions. The dashed cyan curves are the calculated $R_{AA}$ with $g=2.0$ in the absence of the CNM energy loss, while the solid magenta curves are the suppression with $g=1.9$ that includes the CNM energy loss effect. We find that at relatively low $p_T\lesssim 25$ GeV, the two curves are degenerate and yield practically the same suppression. However, at high $p_T\gtrsim 75$ GeV, the calculations with the CNM energy loss effect included  appear to describe the data slightly better, even though  the current experimental uncertainty cannot definitively resolve such a difference. We further illustrate such a ``degeneracy'' for mid-peripheral Pb+Pb collisions in the middle panels of Fig.~\ref{ATLAS-eloss}: both the calculations with $g=2.0\pm 0.1$ without CNM energy loss in the left panel, and the one with $g=1.9\pm 0.1$ but with CNM energy loss in the right panel, describe the ATLAS charged hadron suppression equally well.


Similarly, we compare our calculations to ALICE charged hadron production~\cite{Abelev:2012hxa} in central and mid-peripheral Pb+Pb collisions at the LHC $\sqrt{s_{NN}} = 2.76$ TeV in the top and middle panels of Fig.~\ref{ALICE-eloss}, respectively. The top panels show data sets for both 0-5\% and 5-10\% centrality classes. Similarly, the middle panels show both 30-40\% and 40-50\% centrality classes. These experimental centralities bracket the centralities used for our theoretical calculation and, therefore, we expect that theoretical calculations will fall between the respective centrality classes. This is indeed the case in Fig.~\ref{ALICE-eloss}.  In the mid-peripheral case, the middle panels of Fig.~\ref{ALICE-eloss}, the calculations are still
compatible with the experimental data, though the suppression measured by ALICE is slightly larger than the one seen by ATLAS.

The relative contribution of initial-state CNM energy loss and final-state QGP effects is again illustrated in the bottom panels of Fig.~\ref{ALICE-eloss}. In the transverse momentum range covered by  ALICE measurements it is not possible to
differentiate between the curves that have $g = 2.0$  but do not include CNM energy loss and the curves that have $g = 1.9$ and include CNM energy loss.


Last but not least, we compare our results to CMS measurements of inclusive charged hadron suppression in central and mid-peripheral collisions. As in the comparison with ALICE data, the top panels of Fig.~\ref{CMS-eloss} show both 0-5\% and 5-10\% centrality classes while the middle panels of Fig.~\ref{CMS-eloss} show the 30-40\% and 40-50\% centrality classes. There is a one-to-one correspondence between Fig.~\ref{CMS-eloss} for the CMS comparison to Fig.~\ref{ATLAS-eloss}, for the ATLAS comparison and Fig.~\ref{ALICE-eloss}, for the ALICE comparison.  The comparison with CMS data supports the conclusions drawn from our comparison with the ATLAS and ALICE data sets.


\begin{figure}[t!]
\centering
\includegraphics[width=0.47\textwidth]{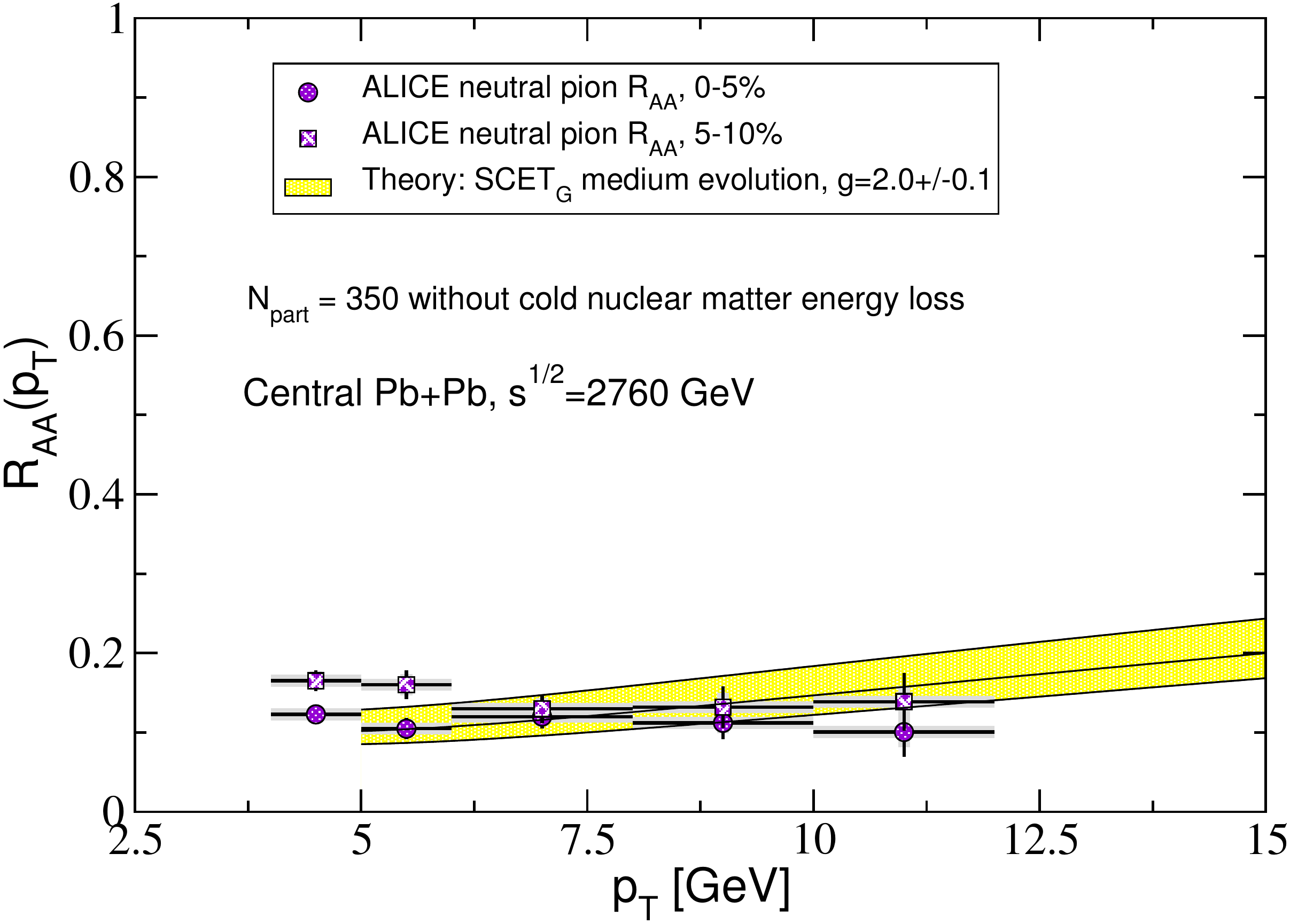}
\hskip 0.2in
\includegraphics[width=0.47\textwidth]{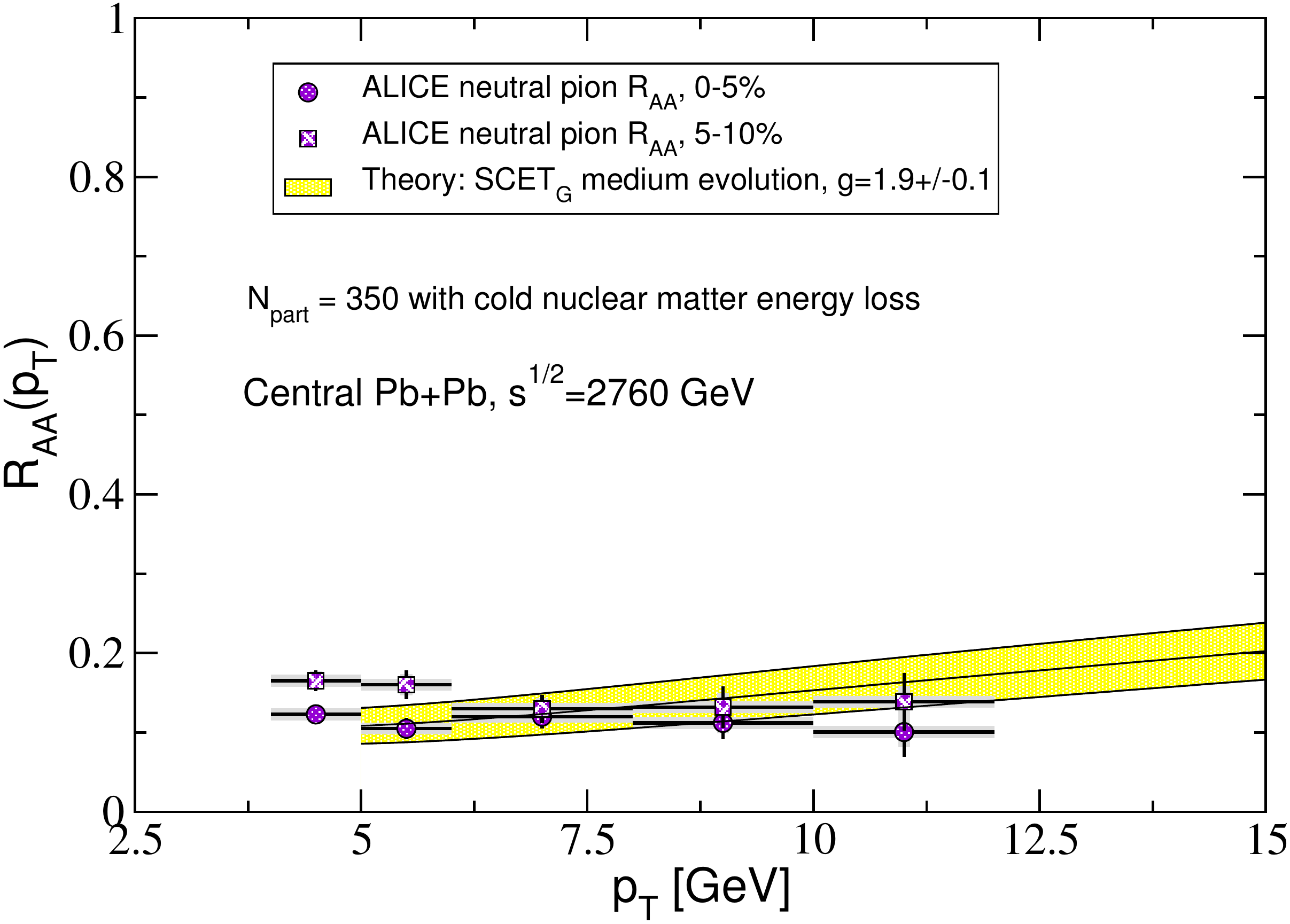}\\
\includegraphics[width=0.47\textwidth]{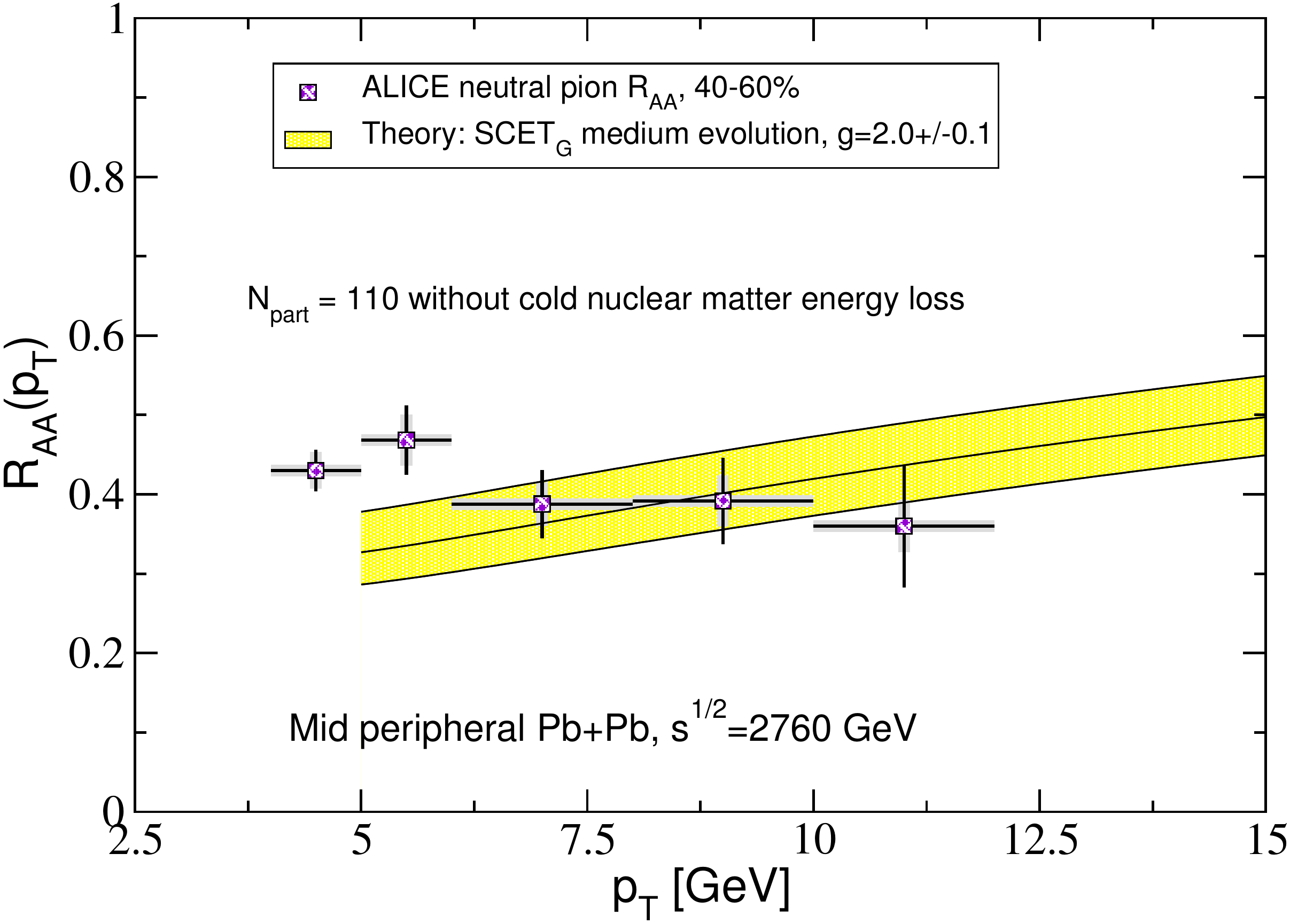}
\hskip 0.2in
\includegraphics[width=0.47\textwidth]{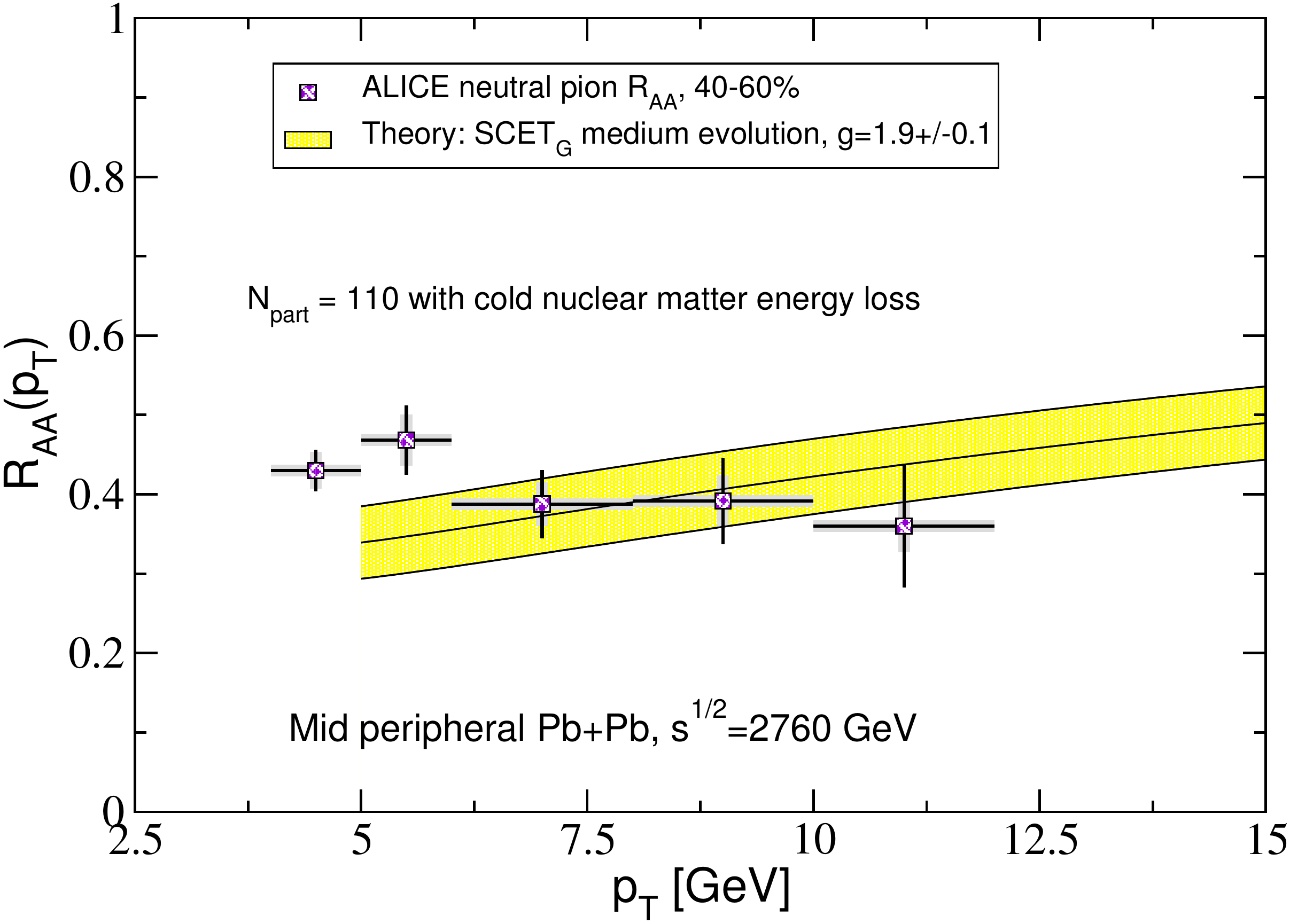}\\
\includegraphics[width=0.47\textwidth]{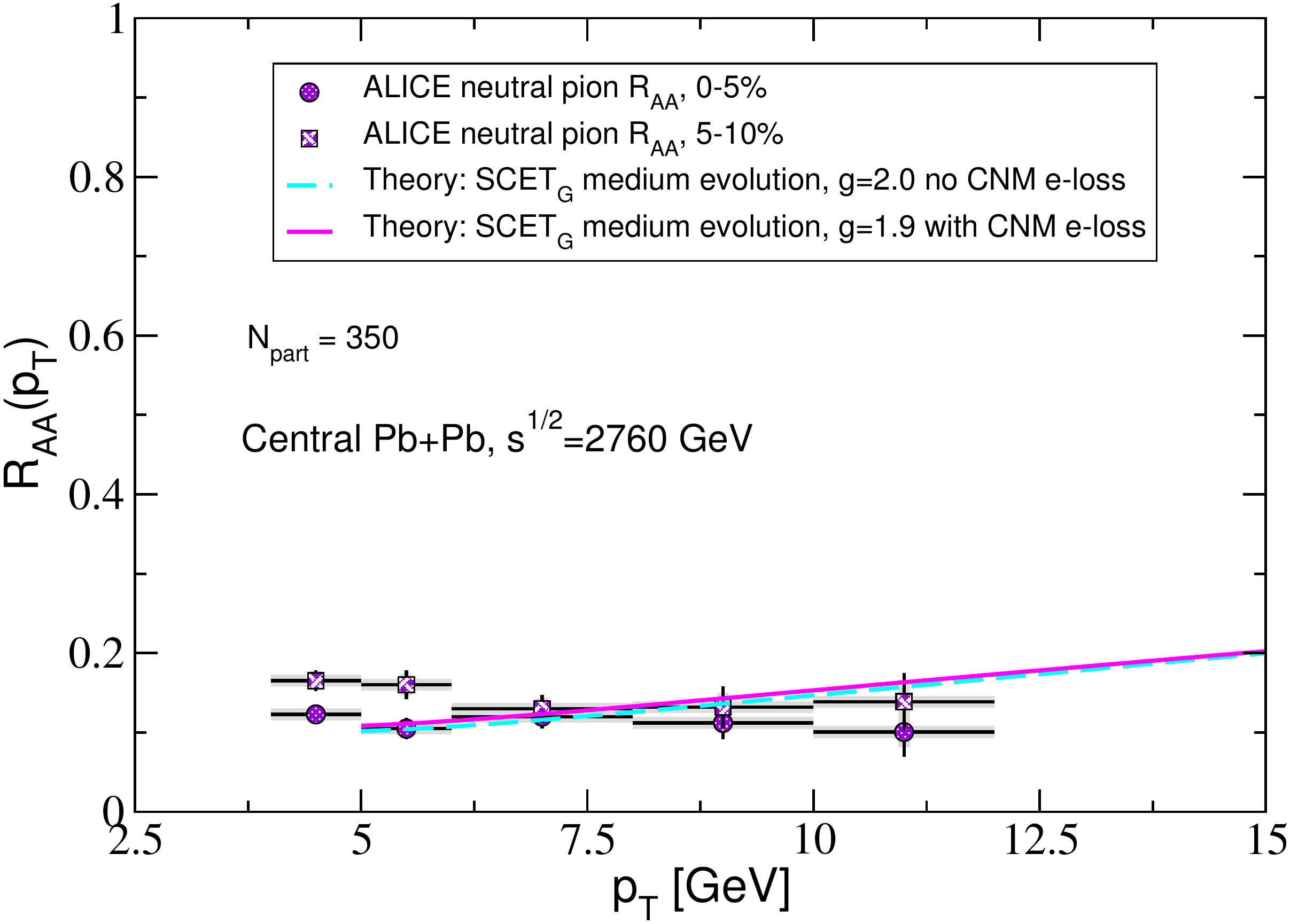}
\hskip 0.2in
\includegraphics[width=0.47\textwidth]{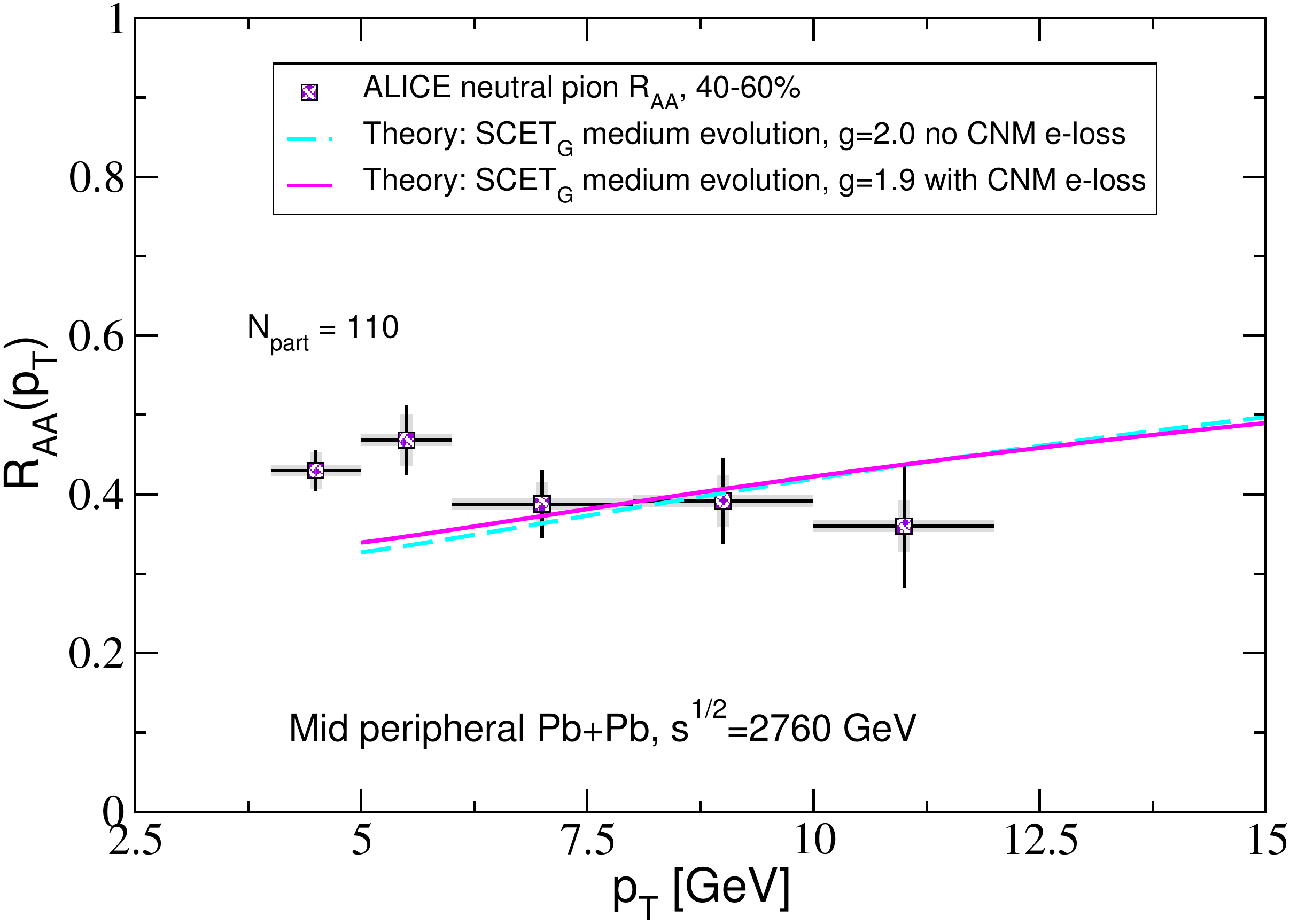}
\caption{Same as in Fig.~\ref{ATLAS-eloss} except the comparison is made to ALICE neutral pions data for the $R_{AA}$ for Pb+Pb collisions at the LHC $\sqrt{s_{NN}}=2.76$ TeV.}
\label{ALICE-pi-eloss}
\end{figure}

We now move on to compare our calculations of $R_{AA}$s for neutral pion production to the ALICE measurements in both central and mid-peripheral Pb+Pb collisions at the LHC in Fig.~\ref{ALICE-pi-eloss}. the top panels of Fig.~\ref{ALICE-pi-eloss} show ALICE data in the 0-5\% and 5-10\% centrality classes, while the middle panels of Fig.~\ref{ALICE-pi-eloss} show data in the 40-60\% centrality class.
Surprisingly, the suppression pattern for neutral pions measured by the ALICE collaboration shows different behavior from the charged hadron measurements: instead of showing decreasing suppression as a function of $p_T$, the neutral pion $R_{AA}$ data from ALICE indicates a constant (or even slightly increasing) suppression over the $p_T$ range. 

There is very little difference in the theoretical calculations for neutral pion production and charged hadron production as seen in all the figures and theoretical predictions for $\sqrt{s_{NN}}=5.1$~TeV given in the next section. This is expected from isospin symmetry $\pi^0 = (\pi^+ + \pi^-)/2$~\cite{Kniehl:2000fe}, and the fact that at high transverse momenta charged pion production dominates the charged hadron multiplicities.  It would be quite useful if additional measurements 
of neutral pion production in Pb+Pb reactions can be performed to higher transverse momenta
for comparison with inclusive charged hadron suppression.

%
%
%

\begin{figure}[t!]
\centering
\includegraphics[width=0.47\textwidth]{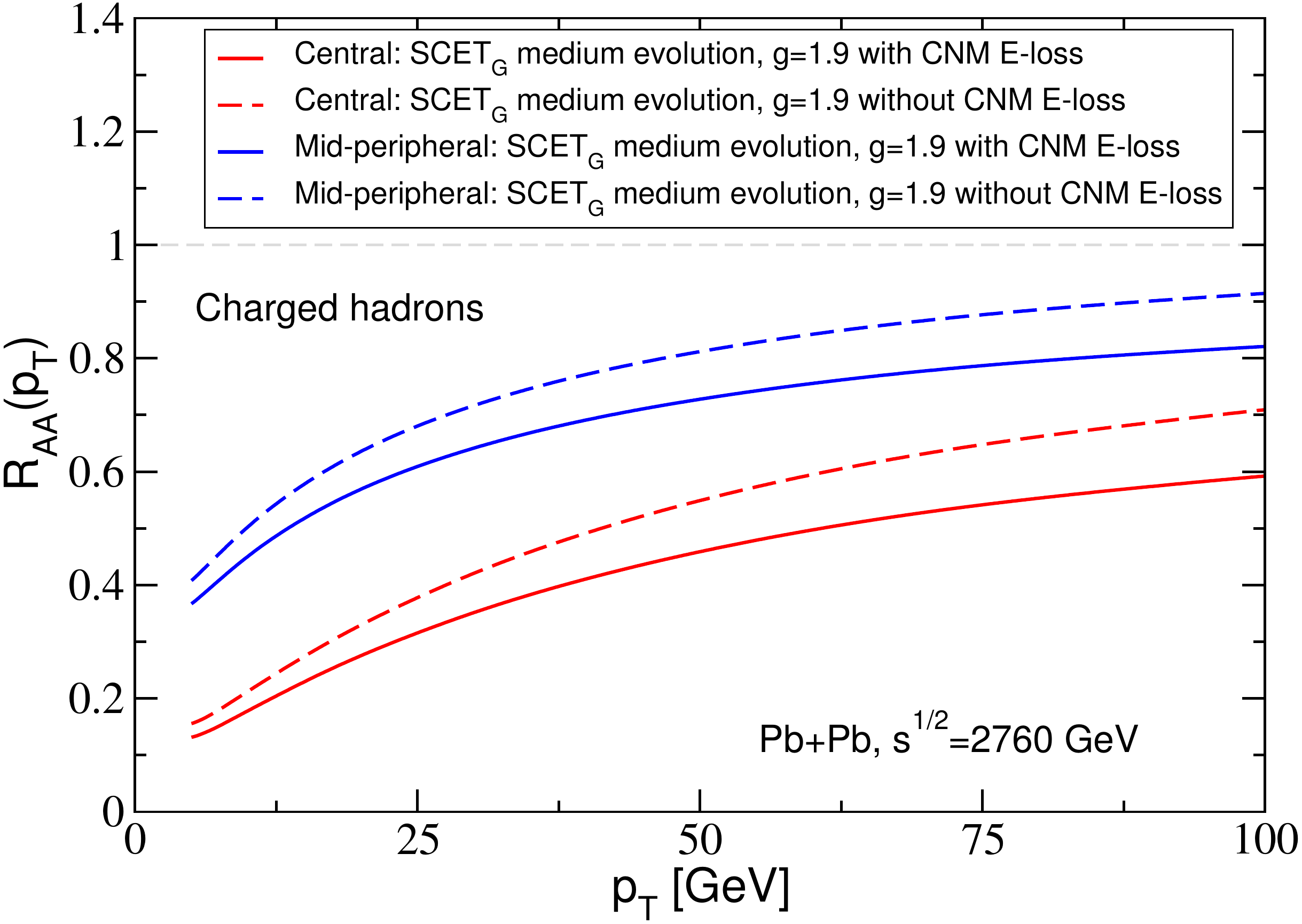}
\hskip 0.2in
\includegraphics[width=0.47\textwidth]{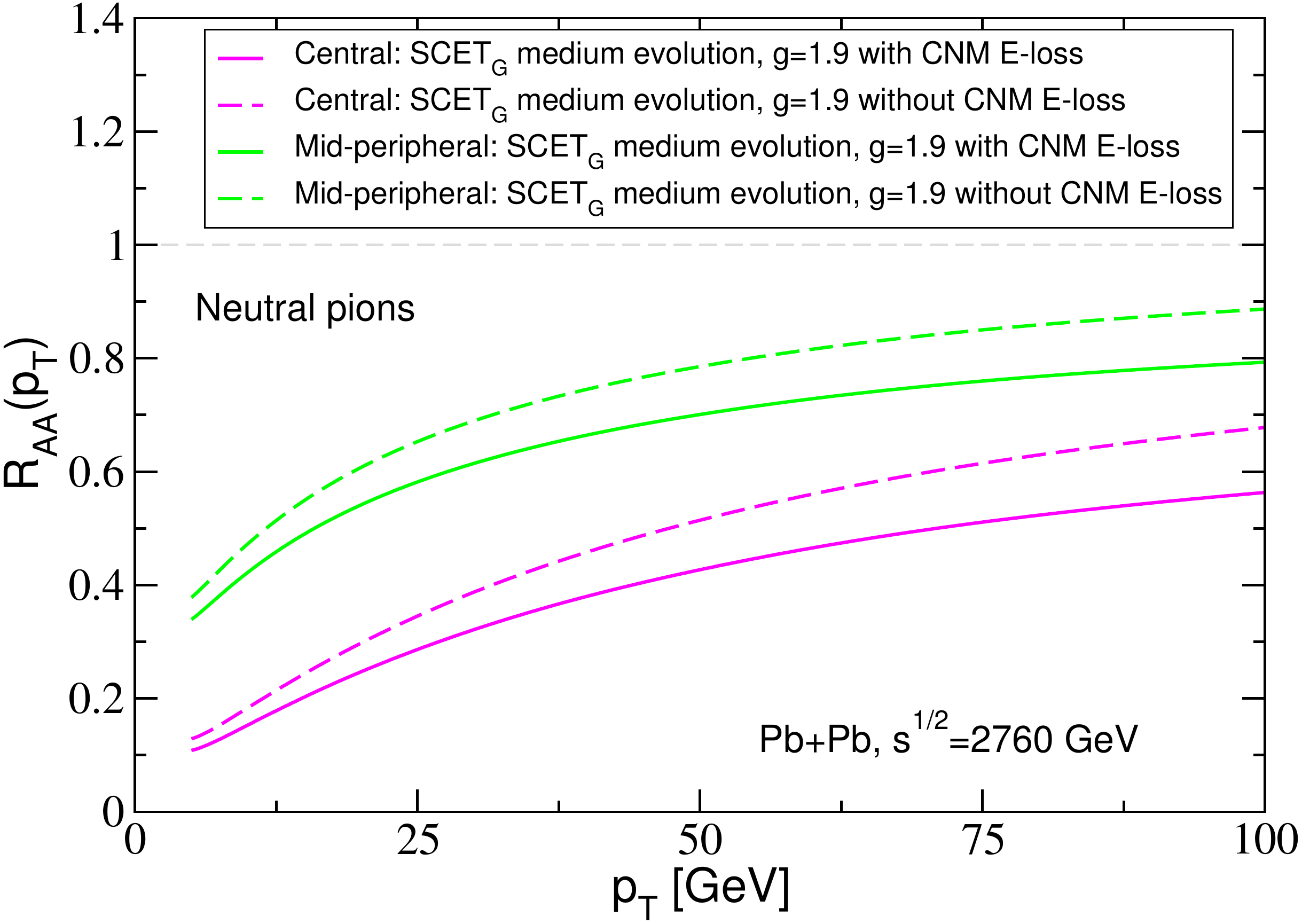}\\
\caption{Comparison of calculations of the nuclear modification factor of charged hadrons (left) and neutral pions (right) with and without cold nuclear matter
energy loss. Results for central and mid-peripheral Pb+Pb collisions are shown for $g=1.9$.}
\label{CNMnoCNM}
\end{figure}

Finally, we note that keeping the coupling constant $g$ fixed and performing calculations with or without cold nuclear matter energy loss will lead to an overall
difference in the magnitude of $R_{AA}$. We illustrate this in Fig.~\ref{CNMnoCNM} for $g=1.9$. In the left panel we show the nuclear modification factor for
light hadrons and  the right panel contains the corresponding  curves for neutral pions.   Solid lines include CNM energy loss and dashed lines do not.

\subsection{Theoretical predictions for the LHC Pb+Pb run II}

The forthcoming Pb+Pb run II at the LHC is anticipated to be performed at center-of-mass energy per nucleon-nucleon pair $\sqrt{s_{NN}} \simeq  5.1$ TeV. 
The higher center-of-mass energy, combined with improvements at the LHC and associated detector experiments, will result in better statistics and data with higher accuracy. In this section, we present  predictions for inclusive charged hadron and neutral pion production in both central and mid-peripheral Pb+Pb collisions.

We will now outline the differences between the calculations of charged hadron and neutral pion suppression factors at $\sqrt{s_{NN}} = 2.76$~TeV  and $ \sqrt{s_{NN}} = 5.1$~TeV.  At higher center-of-mass energies the underlying  spectra of energetic partons are slightly harder in the region of interest, meaning that they have a smaller negative slope.  When convolved with the  fragmentation functions in Eqs.~(\ref{convpp}) and (\ref{convaa}), at higher $\sqrt{s_{NN}}$  smaller momentum fractions $z$ in the fragmentation functions $D_{h/c}(z,Q)$ are sampled, leading to smaller suppression~\cite{Kang:2014xsa}.  
The physics reason for this effect is two-fold. On one hand, the fragmentation functions for light hadrons are falling functions of $z $ and large contribution to the hadronic 
cross section must come from small values of $z$. On the other hand, the partonic cross sections are falling functions of $p_{T_c}$  and large contribution to the hadronic cross section at fixed $p_T$ of the hadron must come from small allowed  values of  $p_{T_c} > p_T$. Since these are competing effects, there is a probabilistic distribution of $z$ around its mean  finite value. When the partonic spectra are harder they allow the $z$ distribution to shift to smaller values. Second, in-medium 
evolution of the fragmentation functions leads to additional softening of $D_{h/c}(z,Q)$ and the effect is most pronounced at large values of $z$, but much less pronounced at smaller values of $z$.

At the same time, for fixed parton or hadron transverse momentum,  smaller   momentum fractions $x_a, \, x_b$ are probed in the parton distribution functions
$f_{a/N}(x_a,Q)$,  $f_{b/N}(x_b,Q)$, leading to an increased fraction of the hard scattered gluons in the final state. 
The strength of the leading diagonal terms in the DGLAP evolution equations   Eqs.~(\ref{eq:AP10}), ~(\ref{eq:AP30}) is proportional to the 
quadratic Casimir in the fundamental (quark) and adjoint (gluon) representations, see Appendix~\ref{medsplitA}.  This will lead to larger in-medium 
softening of  gluon fragmentation functions in comparison to quark fragmentation functions and, consequently, larger  suppression of the cross section
at higher $\sqrt{s_{NN}}$ when the transverse momentum is held fixed.

Finally, the small growth 
of the medium density at larger center-of-mass energies, see Appendix~\ref{numA},  will also lead to larger suppression. 

In terms of cold nuclear matter
energy loss, we anticipate that the additional suppression effect for fixed $p_T$ hadron or jet production will be slightly larger at smaller center-of-mass energies~\cite{Kang:2015mta}. The effect of cold nuclear matter energy loss, see Appendix D,  is larger when $x_a, x_b $ are closer to the kinematic bound. 
We emphasize that  these competing effects are small and we will discuss the net result  for charged hadron and neutral pion quenching below.

\begin{figure}[!t!t]
\centering
\includegraphics[width=0.47\textwidth]{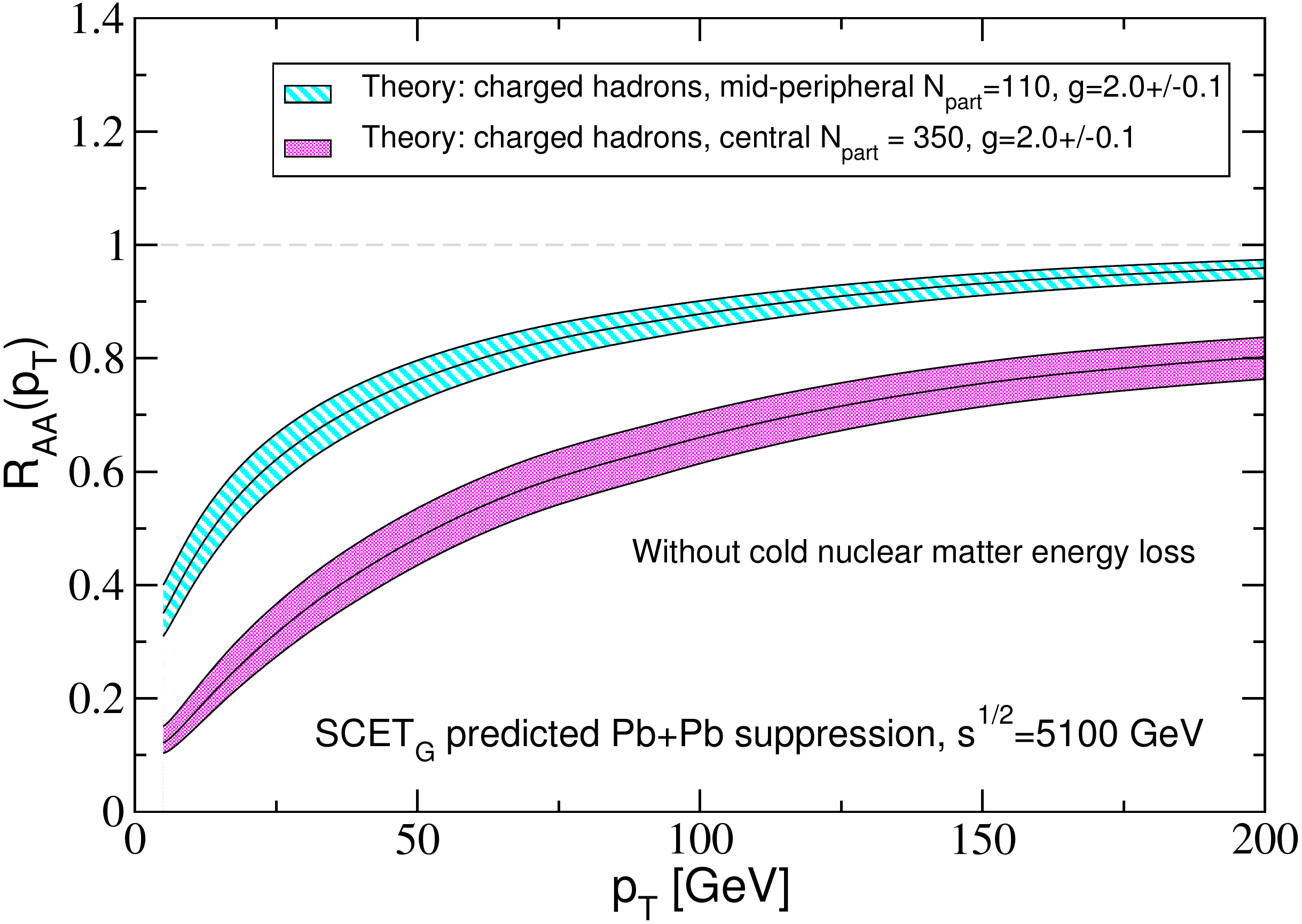}
\hskip 0.2in
\includegraphics[width=0.47\textwidth]{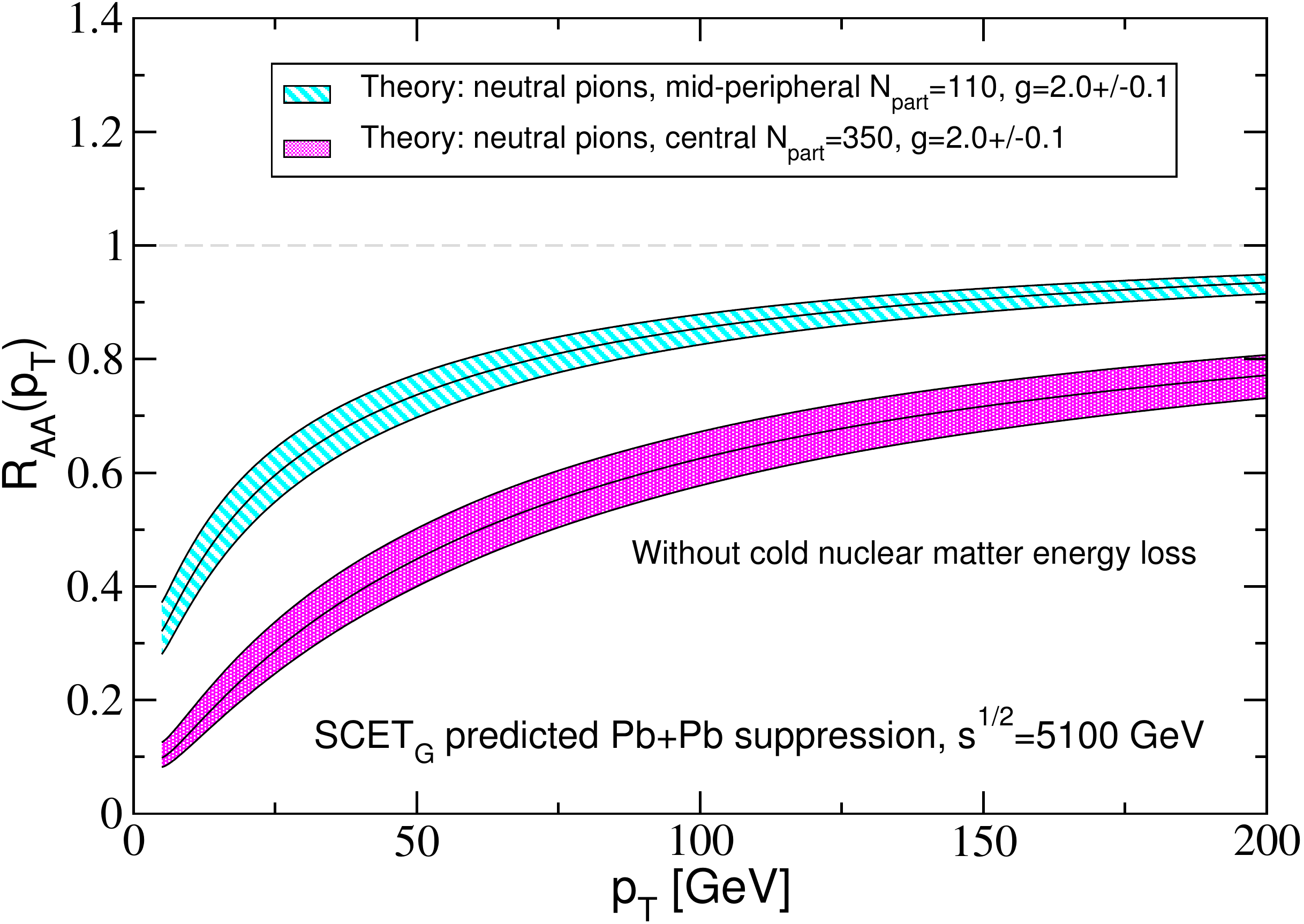}\\
\includegraphics[width=0.47\textwidth]{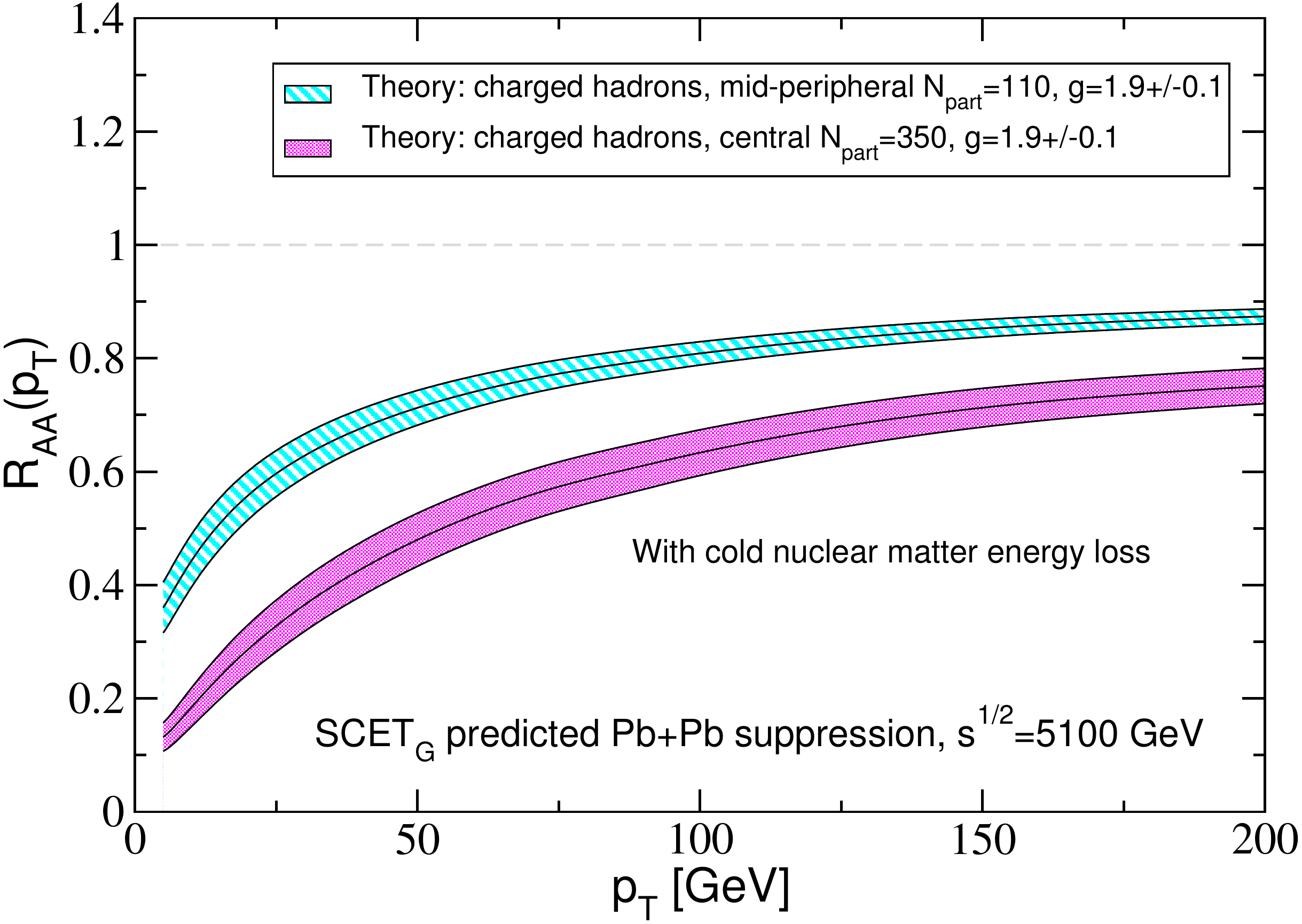}
\hskip 0.2in
\includegraphics[width=0.47\textwidth]{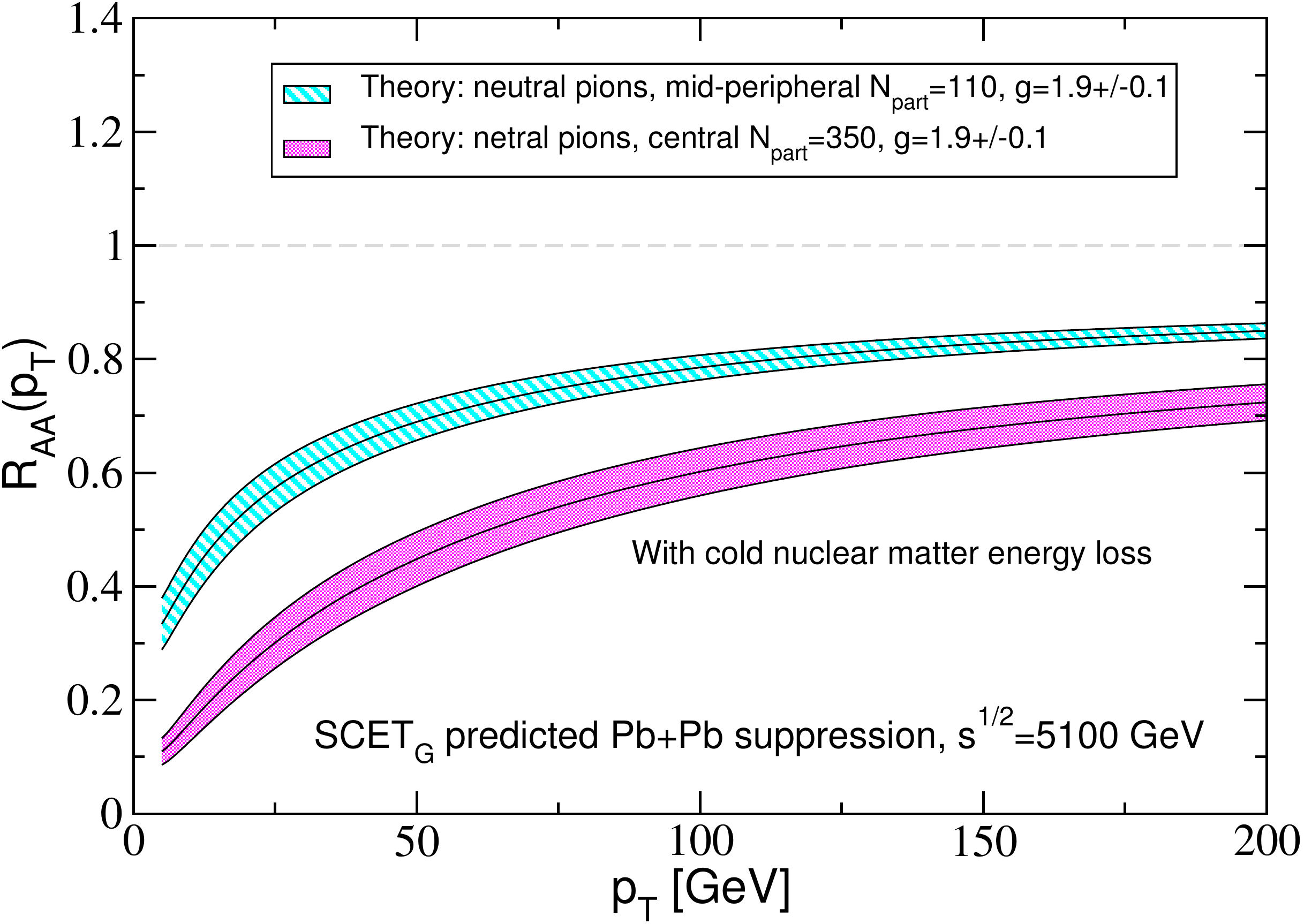}\\
\includegraphics[width=0.47\textwidth]{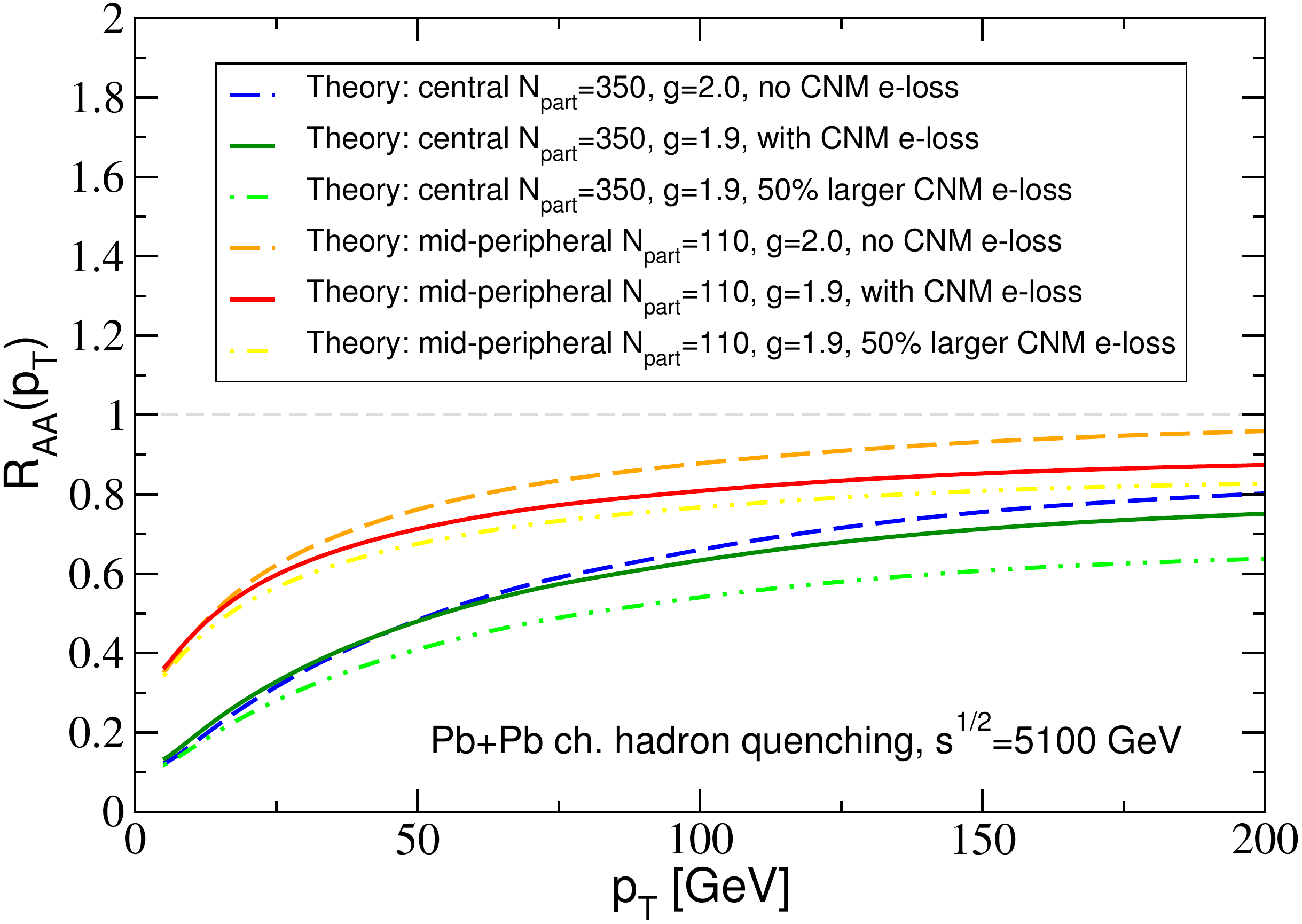}
\hskip 0.2in
\includegraphics[width=0.47\textwidth]{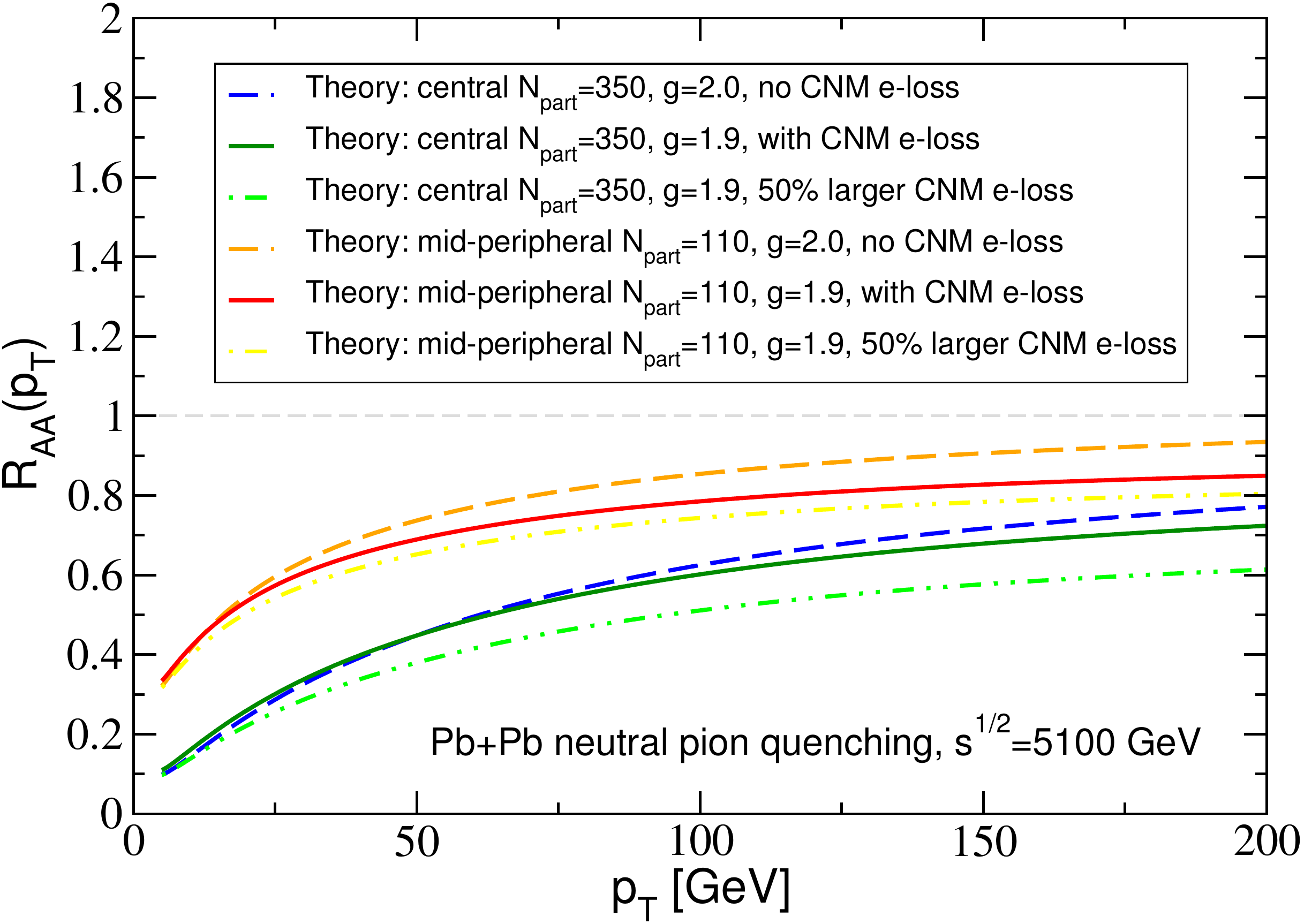}
\caption{The predicted nuclear modification factor $R_{AA}$ for charged hadrons (left) and neutral pions (right) in central (magenta) and mid-peripheral (cyan) Pb+Pb collisions at the LHC $\sqrt{s_{NN}}=5.1$ TeV is plotted as a function of hadron transverse momentum $p_T$. In top panels the bands represent a variation of the coupling strength $g=2.0\pm 0.1$ and we do not include CNM energy loss effects. In the middle panels the coupling is varied $g=1.9\pm 0.1$ and we do include CNM energy loss effects. In the bottom we present the combined results. For the coupling strength $g=2.0$, we again do not include the CNM energy loss effects. For the coupling strength $g=1.9$, we include the usual CNM energy loss constrained at RHIC, as well as an additional calculation where we assume 50\% larger CNM energy loss effects.}
\label{prediction-g1.9}
\end{figure}


We present our predictions for the LHC Pb+Pb run II in Fig.~\ref{prediction-g1.9}. In the top panels of Fig.~\ref{prediction-g1.9} the predicted nuclear modification factor $R_{AA}$ for charged hadrons (left) and neutral pions (right) in central (magenta) and mid-peripheral (cyan) Pb+Pb collisions at the LHC $\sqrt{s_{NN}}=5.1$ TeV are plotted as a function of the hadron transverse momentum $p_T$. The bands represent a variation of the coupling strength $g=2.0\pm 0.1$ and no CNM energy loss effects. The shape of the nuclear modification factor is very similar  to the one calculated and observed in $\sqrt{s_{NN}}=2.76$~TeV reactions.  More specifically, in the case of central Pb+Pb collisions and charged hadron production without CNM energy loss at  $p_T=10$~GeV the predicted charged hadron suppression is 4\% larger  than for LHC run I. At  $p_T=100$~GeV, the predicted suppression difference is less than 1\%.

The middle panels of Fig.~\ref{prediction-g1.9} show a similar calculations to the top ones except that the coupling $g=1.9\pm 0.1$ is used and CNM energy loss effects are included. Compared to the top panels, this results in a slightly slower increase of the $R_{AA}$ with $p_T$ (more quenching). The nuclear modification factor calculations once again are very similar to what was observed at $\sqrt{s_{NN}}=2.76$\,TeV. A more quantitative analysis shows that at $p_T=10$~GeV, the predicted charged hadron suppression is 3\% smaller than
for LHC run I. At  $p_T=100$~GeV, the predicted suppression is 7\% smaller. Our results are consistent with the expectation that CNM energy loss plays 
a more important role at higher transverse momenta.

In the bottom panels of Fig.~\ref{prediction-g1.9} we show in  more detail what role  CNM energy loss may play in inclusive hadron production. We present scenarios: (1) the calculation with $g=2.0$ in absence of CNM energy loss, (2) the calculation with $g=1.9$ and CNM energy loss effects included, and (3) same as (2) but with 50\% larger CNM energy loss. The motivation for discussing this last case is that inclusive jet production in p+A collisions can be compatible with larger energy 
loss~\cite{Kang:2015mta}.   We find that the CNM energy loss effects mainly reduce the slope of the nuclear modification factor $R_{AA}$ as a function of $p_T$, in other words, slow down the rise in  $R_{AA}$ as $p_T$ increases. With better precision, the future measurements at the LHC might be able to resolve the relative contribution from the CNM energy loss and the hot QCD medium effects. It should be noted, though, that the expected differences are relatively small.


\section{Conclusions}
\label{sec:conclusions}

Recent developments in SCET$_{\rm G}$, an effective theory for jet propagation in matter, have  allowed for qualitative advances in understanding 
particle and jet production in heavy ion reactions and quantitative control on the systematic uncertainties that arise from the implementation of final-state inelastic scattering processes in the QGP in jet quenching phenomenology.  A new theoretical framework to describe inclusive particle production and suppression in the heavy-ion environment was laid out in~\cite{Kang:2014xsa}. Here,  we focused on the analytic connection between the
new approach, based upon generalized DGLAP evolution equations  for the fragmentation functions in dense strongly-interacting matter, and the concept of parton energy loss for hard processes.  We found that the traditional energy loss phenomenology is a special soft gluon emission limit of the general QCD evolution framework. With new results for the medium-evolved fragmentation functions at hand,  we combined them with initial-state cold nuclear matter effects for applications to inclusive hadron production in A+A reactions and jet quenching phenomenology.

We compared the nuclear modification factor $R_{AA}$ for charged hadrons measured by ATLAS, ALICE and CMS collaborations at $\sqrt{s_{NN}}=2.76\,\text{TeV}$ to our predictions for central and mid-peripheral collisions. Overall, we found very good qualitative and quantitative agreement, which improves with the inclusion of the cold nuclear matter effects that make the high $p_T$ rise of $R_{AA}$ less steep.
In the case of ALICE, we also compared to the neutral pion data.  We found that even though theoretical predictions and experimental results 
 still mostly agree within the error bars, there are qualitative differences in the shape of $R_{AA}$.  Data appears to show suppression that is flat 
with $p_T$, albeit in a limited transverse momentum range. At high  $p_T$ our theory predicts smaller suppression, essentially identical to that
of charged hadrons. It will be very illuminating to have $\pi^0$   $R_{AA}$s to much higher $p_T$ in the future.  
 From our comparison to data we find that the coupling between the jets in the dense QCD matter created at the LHC is $g=2.0\pm 0.1$ 
in the absence of CNM energy loss and is $g=1.9\pm 0.1$ when CNM energy loss effects are included.
 
For the second Pb+Pb run at the LHC, expected to be at $\sqrt{s_{NN}} \simeq 5.1\text{\,TeV}$, we presented predictions for the $p_T$ dependence of the nuclear modification factors for charged hadrons and neutral pions in central and mid-peripheral collisions. We showed that even though at small and intermediate transverse momenta it is not possible to differentiate between phenomenological results that include CNM energy loss and a weaker coupling between the jet and the medium and  the ones that  do not  include CNM energy loss but have larger coupling between the jet and the medium, such distinction might be possible 
at high $p_T$ with enough statistics.   

Given the need for an accurate theory to describe reconstructed jet production in heavy-ion collisions beyond the energy loss approach, a logical next step will be to  incorporate the full SCET$_{\rm G}$  in-medium splitting functions in the evaluation of jet cross sections~\cite{Chien:2014zna} and jet substructure observables, such as jet shapes~\cite{Chien:2014nsa}.  We leave these developments for future work.

\section*{Acknowledgements}
This work is supported by the US Department of Energy, Office of Science under  Contract Nos. DE-AC52-06NA25396 and DE-SC0011095, by the DOE Early Career Program, as well as in part by the LDRD program at Los Alamos National Laboratory.

\appendix

\section{Medium-induced parton splittings }
\label{medsplitA}

Here, we provide the expressions for the medium-induced splittings for completeness. The calculation to first order in opacity 
 takes into  account the contribution from the splitting induced by the interactions along the trajectory 
of the parent parton and the dominant interference with the splitting induced by the large $Q^2$ process. 
The  topology and kinematics are the same for all the splitting processes. Consequently, all results can be expressed in terms of 
universal transverse momentum vectors  
$\vc{A}_{\perp}, \vc{B}_{\perp}, \vc{C}_{\perp}, \vc{D}_{\perp}$ 
and interference phases $\Omega_1,...,\Omega_5$, defined in~\cite{Ovanesyan:2011xy}:
\begin{eqnarray}
&&\vc{A}_{\perp}=\vc{k}_{\perp},\,\, \vc{B}_{\perp}=\vc{k}_{\perp} + x \vc{q}_{\perp} , \,\, 
\vc{C}_{\perp}=\vc{k}_{\perp} -  (1-x)\vc{q}_{\perp}, \,\,  \vc{D}_{\perp}=\vc{k}_{\perp}-\vc{q}_{\perp},\,\, \\
&&\Omega_1-\Omega_2=\frac{\vc{B}_{\perp}^2}{p_0^+ x(1-x)}, \,\Omega_1-\Omega_3=\frac{\vc{C}_{\perp}^2}{p_0^+x(1-x)},
\,\,  \Omega_2-\Omega_3=\frac{\vc{C}_{\perp}^2-\vc{B}_{\perp}^2}{p_0^+x(1-x)}, \,\,
\Omega_4=\frac{\vc{A}_{\perp}^2}{p_0^+x(1-x)},   \,\,  \Omega_5=\frac{\vc{A}_{\perp}^2-\vc{D}_{\perp}^2}{p_0^+x(1-x)},
\qquad
\end{eqnarray}
where $ p_0^+ = p^+ + k^+ $ and the parent parton has no net transverse momentum. Recall that $x = k^+/p_0^+$
and $\vc{k}_{\perp}$ is the transverse momentum of the parton carrying momentum fraction $x$ relative to the 
parent parton's direction. 

The expressions for the continuous part of the functions read:
\begin{eqnarray}
\left( \frac{dN}{ dxd^2\vc{k}_{\perp} }\right)_{i}=\frac{\alpha_s}{2\pi^2} \left[P^{\text{real}}_{\text{vac}}(x)\right]_i\frac{h_i(x,\vc{k}_{\perp};\beta)}{\vc{k}_{\perp}^2}, \label{eq:hdefinition}
\end{eqnarray}
where $i$ represents one of the four splittings $i=q\rightarrow qg, q\rightarrow gq, g\rightarrow gg, g\rightarrow q\bar{q}$\, and $\left[P^{\text{real}}_{\text{vac}}(x)\right]_i$ is the corresponding vacuum splitting function. Functions $h_i$ equal:
\begin{eqnarray}
h_{q\rightarrow qg}(x,\vc{k}_{\perp};\beta)  &=&  \vc{k}_{\perp}^2  \int \frac{d\Delta z}{\lambda_g(z)}  
\int d^2{\bf q}_\perp  \frac{1}{\sigma_{el}} \frac{d\sigma_{el}^{\; {\rm medium}}}{d^2 {\bf q}_\perp} 
\; \Bigg[  \frac{\vc{B}_{\perp}}{\vc{B}_{\perp}^2} \mcdot \left( \frac{\vc{B}_{\perp}}{\vc{B}_{\perp}^2}  -  \frac{\vc{C}_{\perp}}{\vc{C}_{\perp}^2}   \right) \big( 1-\cos[(\Omega_1 -\Omega_2)\Delta z] \big)
\nonumber \\
& & + \frac{\vc{C}_{\perp}}{\vc{C}_{\perp}^2} \mcdot \left( 2 \frac{\vc{C}_{\perp}}{\vc{C}_{\perp}^2}   
-    \frac{\vc{A}_{\perp}}{\vc{A}_{\perp}^2} - \frac{\vc{B}_{\perp}}{\vc{B}_{\perp}^2}  \right) \big(1- \cos[(\Omega_1 -\Omega_3)\Delta z] \big)
+ \frac{\vc{B}_{\perp}}{\vc{B}_{\perp}^2} \mcdot \frac{\vc{C}_{\perp}}{\vc{C}_{\perp}^2} 
\big( 1 -  \cos[(\Omega_2 -\Omega_3)\Delta z] \big)  
 \nonumber \\  
&&
+ \frac{\vc{A}_{\perp}}{\vc{A}_{\perp}^2} \mcdot \left( \frac{\vc{D}_{\perp}}{\vc{D}_{\perp}^2} - \frac{\vc{A}_{\perp}}{\vc{A}_{\perp}^2} \right) 
\big(1-\cos[\Omega_4\Delta z]\big)  -\frac{\vc{A}_{\perp}}{\vc{A}_{\perp}^2} \mcdot \frac{\vc{D}_{\perp}}{\vc{D}_{\perp}^2}\big(1-\cos[\Omega_5\Delta z]\big)   
\nonumber \\
&&+  \frac{1}{N_c^2}  \frac{\vc{B}_{\perp}}{\vc{B}_{\perp}^2} \mcdot  \left( \frac{\vc{A}_{\perp}}{\vc{A}_{\perp}^2}  -   
\frac{\vc{B}_{\perp}}{\vc{B}_{\perp}^2}      \right)
\big( 1-\cos[(\Omega_1 -\Omega_2)\Delta z] \big)   \Bigg] \, , \\
h_{q\rightarrow gq}(x,\vc{k}_{\perp};\beta)  &=& h_{q\rightarrow qg}(x,\vc{k}_{\perp};\beta) (x\rightarrow 1-x) \, , 
\end{eqnarray}

\begin{eqnarray}
h_{ \left\{ \begin{array}{c}   g \rightarrow gg\\     g\rightarrow q\bar{q}  \end{array} \right\} }(x,\vc{k}_{\perp};\beta)
&= &
\vc{k}_{\perp}^2\int {d\Delta z}   \left\{ \begin{array}{c}    \frac{1}{\lambda_g(z)} \\[1ex]  \frac{1}{\lambda_q(z)}    \end{array} \right\} \int d^2{\bf q}_\perp  \frac{1}{\sigma_{el}} \frac{d\sigma_{el}^{\; {\rm medium}}}{d^2 {\bf q}_\perp} \Bigg[  2\, \frac{\vc{B}_{\perp}}{\vc{B}_{\perp}^2} \mcdot \left(\frac{\vc{B}_{\perp}}{\vc{B}_{\perp}^2}-\frac{\vc{A}_{\perp}}{\vc{A}_{\perp}^2}\right)    
 \big( 1-\cos[(\Omega_1 -\Omega_2)\Delta z]  \big) 
\nonumber \\
& &
+2\, \frac{\vc{C}_{\perp}}{\vc{C}_{\perp}^2} \mcdot 
\left(\frac{\vc{C}_{\perp}}{\vc{C}_{\perp}^2}-\frac{\vc{A}_{\perp}}{\vc{A}_{\perp}^2}\right) 
 \big( 1-\cos[(\Omega_1 -\Omega_3)\Delta z]  \big) +  \left\{ \begin{array}{c}   - \frac{1}{2} \\  \frac{1}{N_c^2-1} \end{array} \right\}
\Bigg(2 \frac{\vc{B}_{\perp}}{\vc{B}_{\perp}^2}\mcdot\left(\frac{\vc{C}_{\perp}}{\vc{C}_{\perp}^2}-\frac{\vc{A}_{\perp}}{\vc{A}_{\perp}^2}\right)
\nonumber \\
 & & \times
\big(1-\cos[(\Omega_1-\Omega_2)\Delta z ]\big)
 +2\,\frac{\vc{C}_{\perp}}{\vc{C}_{\perp}^2}\mcdot\left(\frac{\vc{B}_{\perp}}{\vc{B}_{\perp}^2}-\frac{\vc{A}_{\perp}}{\vc{A}_{\perp}^2}\right)
 \big(1-\cos[(\Omega_1-\Omega_3)\Delta z]\big)
\nonumber \\ 
&& -2\,\frac{\vc{C}_{\perp}}{\vc{C}_{\perp}^2}\mcdot \frac{\vc{B}_{\perp}}{\vc{B}_{\perp}^2}
 \big(1-\cos[(\Omega_2-\Omega_3)\Delta z]\big) +2\,\frac{\vc{A}_{\perp}}{\vc{A}_{\perp}^2}\mcdot\left(\frac{\vc{A}_{\perp}}{\vc{A}_{\perp}^2}-\frac{\vc{D}_{\perp}}{\vc{D}_{\perp}^2}\right)  \big(1-\cos[\Omega_4\Delta z]\big)
\nonumber\\
 && 
+2\,\frac{\vc{A}_{\perp}}{\vc{A}_{\perp}^2}\mcdot \frac{\vc{D}_{\perp}}{\vc{D}_{\perp}^2}
 \big(1-\cos[\Omega_5\Delta z]\big)\Bigg) \Bigg] \, .     
\label{CohRadSX1} 
\end{eqnarray}
Here, $\lambda_q(z)$,  $\lambda_g(z)$ are the scattering lengths of quarks and gluons in the medium and $\left(1/\sigma_{el}\right) \,{d\sigma_{el}^{\; {\rm medium}}}/{d^2 {\bf q}_\perp}$ stands for normalized elastic scattering cross section of a parton in the medium. Even though this quantity varies when the 
parton is a quark or a gluon, in the  high energy limit, when the $t-$ channel dominates the elastic scattering, this normalized cross section does not change significantly. The symmetry of $g\rightarrow gg, g\rightarrow q\bar{q}$ splitting 
kernels under $x \rightarrow 1-x $ is most easily verified explicitly by realizing that the parton scattering 
cross section in the medium is invariant under  $\vc{q}_\perp \rightarrow   -\vc{q}_\perp$.

\section{The soft gluon energy loss limit }
\label{softA}
The energy loss phenomenology, widely used to describe jet quenching at RHIC and the LHC,
is self-consistent only in the soft gluon limit.  It is instructive to verify that in this  small-$x$ limit  only two 
of the four medium-induced splitting intensities survive:     
\begin{eqnarray}
x \left( \frac{dN}{ dx}\right)_{ \left\{ \begin{array}{c}   q \rightarrow qg \\    
 g\rightarrow gg  \\   \end{array} \right\} }  &=&  \frac{\alpha_s}{\pi^2} 
  \left\{ \begin{array}{c}   C_F[ 1+ {\cal O}(x) ] \\  C_A[ 1+ {\cal O}(x) ] 
  \end{array} \right\} 
   \int   \frac{d\Delta z}{\lambda_g(z)}  \int d^2\vc{k}_\perp d^2{\bf q}_\perp  \frac{1}{\sigma_{el}} 
\frac{d\sigma_{el}^{\; {\rm medium}}}{d^2 {\bf q}_\perp} \;
  \nonumber \\
&&  \qquad   \times
 \frac{2 \vc{k}_{\perp}  \cdot \vc{q}_{\perp} }{\vc{k}_{\perp}^2 (\vc{k}_{\perp}-\vc{q}_{\perp})^2}
   \left [ 1-\cos \frac{   (\vc{k}_{\perp}-\vc{q}_{\perp})^2}{xp^+_0} \Delta z \right].
\label{smallx}
\end{eqnarray} 
Note that the fractional intensity is the medium-induced splitting kernel weighed by the 
large lightcone momentum fraction of the emitted soft gluon and integrated over the 
available transverse momentum space. The remaining two splitting intensities  $g \rightarrow q\bar{q}$,  $q \rightarrow gq$
are suppressed in this limit by a power of the $x\ll1$, Specifically, they go as  
 $ T_R [{\cal O}(x/2)]  $,  $ C_F [ {\cal O}(x/2)] $.
In this limit the interference structure for all medium induced splitting functions is the same. \\

Factoring the vacuum splitting function in the soft gluon approximation out according to \eq{eq:hdefinition} and using the small $x$ limit of the $\vc{k}_\perp$ unintegrated medium-induced intensities in \eq{smallx} we obtain for both quark and gluon splittings identical function $h^{\text{sga}}$:
\begin{eqnarray}
h^{\text{sga}}(x,\vc{k}_\perp;\beta)=\int_0^L\frac{\d\Delta z}{\lambda_g}\,\d^2\vc{q}_{\perp}\frac{1}{\sigma_{el}}\frac{\d\sigma_{el}^{\text{medium}}}{\d^2\vc{q}_{\perp}}\frac{2\,\vc{k}_{\perp}\mcdot \vc{q}_{\perp}}{(\vc{k}_{\perp}-\vc{q}_{\perp})^2}\left[1-\cos\frac{(\vc{k}_{\perp}-\vc{q}_{\perp})^2}{xp_0^+}\Delta z\right].
\end{eqnarray}
The medium-induced splitting functions in the soft gluon approximation are given in terms of this function in Eqs.~(\ref{eq:Pqqsmallx}, \ref{eq:Pggsmallx}) above.

\section{Numerical methods and  medium  properties}
\label{numA}
Even though this paper is focused on understanding the similarities and differences between the energy loss 
approach and the treatment of medium-induced parton splittings on the same footing as the 
vacuum  branchings, we pay special attention to the numerical approach and the results are 
applicable to phenomenology.

We implement the nuclear geometry via a standard optical Glauber model. The inelastic proton-proton
scattering cross section can be obtained from the Particle Data Group~\cite{PDG} or directly from
measurements. For example, the value we use at $\sqrt{s_{NN}}=2.76$~TeV is $\sigma_{in} = 64$~mb, well within the
ALICE $1\sigma$ result $62.8^{+2.8}_{-4.0} \pm 1.2$~mb~\cite{Abelev:2012sea}.   For  $\sqrt{s_{NN}}=5.1$~TeV 
we use the value $\sigma_{in} = 70$~mb.  In the plane transverse to the collision axis the hard scattering
processes are distributed according to the binary collision density  $ \sim d^2N_{\rm coll}/d^2{\bf x}_\perp $.
Results relevant to the LHC phenomenology  are calculated using full  numerical evaluation  of the 
medium-induced splitting functions.  
In contrast to jet production, the medium is distributed according to the number of participants
density  $ \sim d^2N_{\rm part}/d^2{\bf x}_\perp $. Soft particles that carry
practically all of the energy deposited in the  heavy ion collision cannot  deviate
a lot from such scaling.  We take into account longitudinal
Bjorken expansion since transverse expansion leads to noticeable
corrections only in the extreme transverse velocity $\beta_T \rightarrow 1$
limit~\cite{Gyulassy:2001kr}. In our approach all relevant
finite time   and finite kinematics integrals, such as the ones over the separation   between
the scattering centers $\Delta z_i = z_i - z_{i-1}$, the bremsstrahlung
gluon phase space ($k^+, \vc{k}_\perp$) and the Glauber gluon transverse momentum 
${\bf q}_\perp$ distributions are done numerically.

The evolving intrinsic momentum and length scales in the QGP expected to
be created at the LHC are determined as follows:  we first estimate the
QGP formation times at RHIC and the LHC to be $\tau_0 =  0.5 $~fm  and $\tau_0 =  0.3 $~fm, respectively.  Gluons
dominate the soft parton multiplicities at the LHC and their time- and
position-dependent density can be related to charged hadron rapidity density in the Bjorken
expansion model~\cite{Markert:2008jc}:
\begin{equation}
\rho =  \frac{1}{\tau}  \frac{d^2 ( dN_g/dy)}{d^2 {\bf x}_\perp}
\approx  \frac{1}{ \tau } \frac{3}{2}
\left| \frac{d\eta}{dy} \right| \frac{d^2 (dN^{ch}/d\eta) }{d^2 {\bf
x}_\perp}
\;  .
\label{ydep}
\end{equation}
Here,  $dN^{ch}/d\eta = \kappa N_{\rm part}/ 2$ where for central Pb+Pb collisions $\kappa =
8.25 $ at  $\sqrt{s} = 2.76$~TeV and $\kappa = 8.6 $ at $\sqrt{s} = 5.1$~TeV. 
Assuming  local thermal equilibrium one finds:
\begin{equation}
T(\tau, {\bf x}_{\perp}) = \  ^3\!\sqrt{ \pi^2 \rho(\tau,{\bf
x}_{\perp}) / 16 \zeta(3) }\;,  ~ \tau > \tau_0 \; .
\label{tempdet}
\end{equation}
The Debye screening scale is given by $m_D = gT$, recalling that we work
in the
approximation of a gluon-dominated plasma and $N_f = 0$. The relevant
gluon  mean free path is easily evaluated:  $\lambda_g = 1/ \sigma^{gg} \rho$
with
$\sigma^{gg} = (9/2) \pi \alpha_s^2 / m_D^2$.
 In our evaluation we use  an effective coupling between the jet and the medium, for example  $g  =
2$ corresponds to $\alpha_s = 0.32$,  but the QGP-induced bremsstrahlung is calculated with  a
running coupling $ \alpha_s(Q)$ for the emission vertex.  We finally note that in the medium partons acquire
effective thermal mass and we include it as follows $\vc{k}^2_\perp \rightarrow \vc{k}^2_\perp +m_D^2$ 
in the non-asymptotic temperature regime.

\section{Implementation of cold nuclear matter effects }
\label{softD}

To leading order in the framework of factorized perturbative QCD, in p+p collisions, 
leading parton production can be written as follows:
\begin{eqnarray}
 \frac{d \hat \sigma_c }{dyd^2 p_{T_c}}
&=& K\frac{\alpha_s^2}{s}\sum_{a,b,d}\int \frac{dx_a}{x_a}d^2k_{aT} \, 
f_{a/N}(x_a, k_{aT}^2)\int \frac{dx_b}{x_b}d^2k_{bT} \, 
f_{b/N}(x_b, k_{bT}^2)  H_{ab\to cd}(\hat s,\hat t,
\hat u)\delta(\hat s+\hat t+\hat u).
\end{eqnarray}
Here, $s$ is the center-of-mass energy squared and $H_{ab\to cd}(\hat s,\hat t,\hat u)$ are the hard-scattering coefficient 
functions with $\hat s,\hat t,\hat u$ being the usual partonic Mandelstam variables.
A phenomenological $K$-factor can be included to account for higher order QCD 
contributions, but it cancels in the calculations of the nuclear modification factor. $f_{a,b/N}(x, k_{T}^2)$ are the parton distribution functions with 
longitudinal momentum fraction $x$ and transverse component $k_T$. We have included this $k_T$-dependence in 
order to incorporate the Cronin effect in A+A collisions. We assume a Gaussian form in this 
variable:
\begin{equation}
f_{a/N}(x_a, k_{aT}^2) = f_{a/N}(x_a) \frac{1}{\pi\langle k_T^2\rangle} e^{-k_{aT}^2/\langle k_T^2\rangle}, \quad
f_{b/N}(x_b, k_{bT}^2) = f_{b/N}(x_a) \frac{1}{\pi\langle k_T^2\rangle} e^{-k_{bT}^2/\langle k_T^2\rangle},
\label{gauss}
\end{equation}
where $f_{a/N}(x_a)$, $f_{b/N}(x_b)$ are the usual collinear PDFs in a nucleon.

In A+A collisions we take into account the following cold nuclear matter effects:

{\it Isospin effect.} It can be easily accounted for on average in the nPDFs for a 
nucleus with atomic mass $A$ and $Z$ protons via:
\begin{equation}
f_{a,b/A}(x) = \frac{Z}{A} f_{a,b/p}(x) + \left(1-\frac{Z}{A}\right)f_{a,b/n}(x).
\label{iso}
\end{equation}
In Eq.~(\ref{iso}) $f_{a/p}(x)$ and $f_{a/n}(x)$ are the PDFs inside a proton and neutron, respectively.
The PDFs in the neutron are related to the PDFs in the proton via isospin symmetry. 

{\it Cronin effect.}  It can be modeled via multiple initial-state scatterings of the partons 
in cold nuclei and the corresponding induced parton transverse momentum broadening~\cite{Ovanesyan:2011xy}.
If the parton distribution function $f_{a,b/A}(x_{a,b}, k_{a,b,T}^2)$ has a normalized Gaussian form, 
elastic scattering induces further $k_T$-broadening in the nucleus as the parton traverses $L_{eff}$:
\begin{equation}
\langle k_{a, T}^2\rangle_{AA} = \langle k_{a, T}^2\rangle_{pp} 
+ \left\langle \frac{2\mu^2 L_{eff}}{\lambda_{a}}\right\rangle \zeta ,  
 \quad  \langle k_{b, T}^2\rangle_{AA} = \langle k_{b, T}^2\rangle_{pp} 
+ \left\langle \frac{2\mu^2 L_{eff}}{\lambda_{b}}\right\rangle \zeta
\end{equation}
Here $k_{b,T}$ is the transverse momentum component for the parton prior to the hard scattering,
$\zeta=\ln(1+\delta p_T^2)$~\cite{Vitev:2003xu,Vitev:2002pf}, and we choose 
$\delta = 0.3$ GeV$^{-2}$,   $\mu^2$ sets the soft transverse momentum transfer squared and $\lambda_g= C_F/C_A \lambda_q$ are the related gluon and quark scattering lengths to leading order.

{\it Dynamical shadowing.} Power-suppressed resummed coherent final-sate scattering of the struck partons 
leads to shadowing effects (suppression of the cross section in the small-$x$ 
region)~\cite{Qiu:2004da}. The effect can be interpreted as a generation of dynamical parton mass
in the background gluon field of the nucleus~\cite{Qiu:2004qk}. It is included via:
\begin{equation}
x_a\to x_a \left(1+ \frac{\xi^2_c(A^{1/3}-1)}{-\hat u}\right),  \quad x_b\to x_b \left(1+ \frac{\xi_d^2(A^{1/3}-1)}{-\hat t}\right),
\end{equation}
where $x_a, x_b$ is the parton momentum fractions inside the target nuclei and $c,\; d$
are the outgoing partons that rescatter.  $\xi^2$ represents a characteristic 
scale of the multiple scattering  and
$\xi^2_q = C_F/C_A \xi^2_g$~\cite{Qiu:2004da}.

{\it Cold nuclear matter energy loss.} As the partons from one nucleus undergo multiple scattering in 
the other nucleus before the large $Q^2$ process, they can lose energy due to medium-induced gluon bremsstrahlung. 
This effect can be easily implemented as a momentum fraction shift in the PDFs
\begin{equation}
f_{q,g/N}(x_a) \to f_{q,g/N}\left(\frac{x_a}{1-\epsilon_{\rm eff.\, q,g}}\right),
\quad
f_{q,g/N}(x_b) \to f_{g,g/N}\left(\frac{x_b}{1-\epsilon_{\rm eff.\, q,g}}\right).
\label{eloss}
\end{equation}
Ideally,  Eq.~(\ref{eloss}) should include a convolution over the probability distribution
of cold nuclear matter energy loss~\cite{Neufeld:2010dz}. However, concurrent implementation of
such distribution together with the Cronin effect and coherent power corrections is computationally
demanding. The main effect of the fluctuations due to multiple gluon emission is an effective reduced
fractional energy loss  $ \epsilon_{\rm eff.} \approx 0.7 \langle \Delta E/ E \rangle $.  
We evaluate  $\langle \Delta E/ E \rangle$ as in~\cite{Vitev:2007ve} and it also depends on 
the medium parameters $\mu^2$ and $\lambda_g$.  Work is underway to better understand the 
physics of initial-state parton showers at lower fixed target energies versus coherent energy loss at
much higher collider energies~\cite{Ovanesyan:2015dop}. 

We expect that, at least as an approximation, cold nuclear matter parameters that enter the 
evaluation of cold nuclear matter effects are related. For quarks, the scale of coherent scattering 
per nucleon was determined to be  $\xi^2_q = {\cal O}(0.1 \, {\rm GeV}^2) $. 
For the default cold nuclear matter effects used 
in this paper we set $\xi^2_q = 2 \mu^2 / \lambda_q = 0.08$~GeV$^2$ and for the 50\% larger
nuclear effects  $\xi^2_q = 2 \mu^2 / \lambda_q = 0.12$~GeV$^2$. 
Inclusion of these cold nuclear matter effects gives ${d \hat \sigma_c^{\rm CNM} (p_{T_c})} { \big /} {dyd^2 p_{T_c}} $.

\bibliographystyle{h-physrev}
\bibliography{bibliography}
\end{document}